\documentclass[10pt]{article}
\usepackage{amssymb,amsmath}
\usepackage{amsthm}
\usepackage{mathrsfs}
\usepackage{dcolumn}
\usepackage{bm}
\usepackage{fullpage}
\usepackage{color}
\usepackage[all]{xy}
\usepackage[pdftex]{graphicx,hyperref}  
\usepackage{ulem}
\begin{document}
\newtheorem{Definition}{Definition}[section]
\newtheorem{Theorem}{Theorem}[section]
\newtheorem{Proposition}{Proposition}[section]
\newtheorem{Lemma}{Lemma}[section]
\theoremstyle{definition}
\newtheorem*{Proof}{Proof}
\newtheorem{Example}{Example}[section] 
\newtheorem{Postulate}{Postulate}[section]
\newtheorem{Corollary}{Corollary}[section]
\newtheorem{Remark}{Remark}[section]
\newtheorem{Claim}{Claim}[section]

\theoremstyle{remark}
\newcommand{\beq}{\begin{equation}}
\newcommand{\beqa}{\begin{eqnarray}}
\newcommand{\eeq}{\end{equation}}
\newcommand{\eeqa}{\end{eqnarray}}
\newcommand{\non}{\nonumber}
\newcommand{\lb}{\label}
\newcommand{\fr}[1]{(\ref{#1})}
\newcommand{\bb}{\mbox{\boldmath {$b$}}}
\newcommand{\bbe}{\mbox{\boldmath {$e$}}}
\newcommand{\bt}{\mbox{\boldmath {$t$}}}
\newcommand{\bn}{\mbox{\boldmath {$n$}}}
\newcommand{\br}{\mbox{\boldmath {$r$}}}
\newcommand{\bC}{\mbox{\boldmath {$C$}}}
\newcommand{\bp}{\mbox{\boldmath {$p$}}}
\newcommand{\bx}{\mbox{\boldmath {$x$}}}
\newcommand{\bF}{\mbox{\boldmath {$F$}}}
\newcommand{\bT}{\mbox{\boldmath {$T$}}}
\newcommand{\bQ}{\mbox{\boldmath {$Q$}}}
\newcommand{\bS}{\mbox{\boldmath {$S$}}}
\newcommand{\balpha}{\mbox{\boldmath {$\alpha$}}}
\newcommand{\bomega}{\mbox{\boldmath {$\omega$}}}
\newcommand{\ve}{{\varepsilon}}
\newcommand{\e}{\mathrm{e}}
\newcommand{\met}{\mathrm{met}}
\newcommand{\B}{\mathrm{B}}
\newcommand{\E}{\mathrm{E}}
\newcommand{\R}{\mathrm{R}}
\newcommand{\Z}{\mathrm{Z}}
\newcommand{\HH}{\mathrm{H}}
\newcommand{\I}{\mathrm{I}}
\newcommand{\II}{\mathrm{II}}
\newcommand{\hF}{\widehat F}
\newcommand{\hL}{\widehat L}
\newcommand{\tA}{\widetilde A}
\newcommand{\tB}{\widetilde B}
\newcommand{\tC}{\widetilde C}
\newcommand{\tL}{\widetilde L}
\newcommand{\tK}{\widetilde K}
\newcommand{\tX}{\widetilde X}
\newcommand{\tY}{\widetilde Y}
\newcommand{\tU}{\widetilde U}
\newcommand{\tZ}{\widetilde Z}
\newcommand{\talpha}{\widetilde \alpha}
\newcommand{\te}{\widetilde e}
\newcommand{\tv}{\widetilde v}
\newcommand{\ts}{\widetilde s}
\newcommand{\tx}{\widetilde x}
\newcommand{\ty}{\widetilde y}
\newcommand{\ud}{\underline{\delta}}
\newcommand{\uD}{\underline{\Delta}}
\newcommand{\chN}{\check{N}}
\newcommand{\cA}{{\cal A}}
\newcommand{\cC}{{\cal C}}
\newcommand{\cD}{{\cal D}}
\newcommand{\cF}{{\cal F}}
\newcommand{\cH}{{\cal H}}
\newcommand{\cI}{{\cal I}}
\newcommand{\cK}{{\cal K}}
\newcommand{\cL}{{\cal L}}
\newcommand{\cM}{{\cal M}}
\newcommand{\cO}{{\cal O}}
\newcommand{\cQ}{{\cal Q}}
\newcommand{\cS}{{\cal S}}
\newcommand{\cY}{{\cal Y}}
\newcommand{\cU}{{\cal U}}
\newcommand{\cV}{{\cal V}}
\newcommand{\tcA}{\widetilde{\cal A}}
\newcommand{\DD}{{\cal D}}
\newcommand\TYPE[3]{ \underset {(#1)}{\overset{{#3}}{#2}}  }
\newcommand{\Qc}{\overset{\footnotesize\circ}{Q}}
\newcommand{\bfe}{\boldsymbol e} 
\newcommand{\bfb}{{\boldsymbol b}}
\newcommand{\bfd}{{\boldsymbol d}}
\newcommand{\bfh}{{\boldsymbol h}}
\newcommand{\bfj}{{\boldsymbol j}}
\newcommand{\bfn}{{\boldsymbol n}}
\newcommand{\bfA}{{\boldsymbol A}}
\newcommand{\bfB}{{\boldsymbol B}}
\newcommand{\bfJ}{{\boldsymbol J}}
\newcommand{\bfS}{{\boldsymbol S}}
\newcommand{\dr}{\mathrm{d}}
\newcommand{\saddle}{\mathrm{saddle}}
\newcommand{\can}{\mathrm{can}}
\newcommand{\wt}[1]{\widetilde{#1}}
\newcommand{\wh}[1]{\widehat{#1}}
\newcommand{\ch}[1]{\check{#1}}
\newcommand{\ol}[1]{\overline{#1}}
\newcommand{\ii}{\imath}
\newcommand{\ic}{\iota}
\newcommand{\mbbP}{\mathbb{P}}
\newcommand{\mbbR}{\mathbb{R}}
\newcommand{\mbbN}{\mathbb{N}}
\newcommand{\mbbZ}{\mathbb{Z}}
\newcommand{\Leftrightup}[1]{\overset{\mathrm{#1}}{\Longleftrightarrow}}
\newcommand{\avg}[1]{\left\langle\,{#1}\, \right\rangle}
\newcommand{\step}{\lrcorner\hspace*{-0.55mm}\ulcorner}

\title{
  Nonequilibrium thermodynamic
  process with hysteresis and metastable states 
  --- A contact Hamiltonian with unstable and stable segments of a Legendre
  submanifold   
}
\author{ Shin-itiro GOTO 
  \\
  Center for Mathematical Science and Artificial Intelligence,\\
  Chubu University,\quad 
1200 Matsumoto-cho, Kasugai, Aichi 487-8501, Japan
}
%
\date{\today}
\maketitle
\begin{abstract}%
  In this paper,
  a dynamical process in a statistical thermodynamic system of spins exhibiting
  a phase transition is described on a contact manifold, 
  where such a dynamical process is a process that a metastable equilibrium 
  state evolves into the most stable symmetry broken equilibrium state.
  Metastable and the most stable equilibrium states in the
  symmetry broken phase
  or ordered phase are assumed to be described    
  as pruned projections of 
  Legendre submanifolds of contact manifolds,
  where these pruned projections of the submanifolds express hysteresis
  and pseudo-free energy curves.  
  Singularities associated with phase transitions are naturally arose
  in this framework as has been suggested by Legendre singularity theory. 
  Then a particular contact Hamiltonian vector field is proposed so that 
  a pruned segment of the projected Legendre submanifold is a stable fixed
  point set in a region of a contact manifold, and that another
  pruned segment is a unstable fixed point set. 
  This contact Hamiltonian vector field is identified with 
  a dynamical process 
  departing from a metastable equilibrium state to 
  the most stable equilibrium one.  
  To show the statements above explicitly an Ising type spin model with
  long-range interactions, called the Husimi-Temperley model, 
  is focused, where this model exhibits a phase transition.
\end{abstract}%

%
\section{Introduction}
Contact geometry is known as an odd-dimensional analogue of symplectic
geometry\,\cite{Libermann1987,Silva2008,McInerney2013}, and has been
studied from viewpoints of pure and applied mathematics\,\cite{Arnold}. 
From the pure mathematics side,
contact topology\cite{Geiges2008},
contact Riemannian geometry\,\cite{Blair1976},
and so on\,\cite{Wang2019} are studied.
From the applied mathematics side,
geometrization of thermodynamics\,\cite{Harmann1973,Mrugala2000},
statistical mechanics\,\cite{Mrugala1990PRA,Kushner2020}, 
applications to
information geometry\,\cite{Goto2015JMP,Mori2018,Nakajima2021}, 
and so on\,\cite{Ghrist2000,Ohsawa2015,Goto2016,Bravetti2019Celestial,GH2021}
are studied. 
In particular contact geometric approaches to thermodynamics\,
\cite{Schaft2018entropy,Lopez2021} 
and dissipative
mechanics\,\cite{Carinera2019,Bravetti2017,Leon2019} are intensively studied.  
Although the both sides are close in some sense\,\cite{Entov2021}, we feel that
some gap between them exists, and that profound theorems found in
pure mathematics should be applied to such application areas.
By filling such a gap, undiscovered notions and facts 
are expected to be found 
  as in other previous
  contacts between physics and geometry\,\cite{Frenkel,Nakahara}.

Nonequilibrium statistical mechanics and thermodynamics are developing 
branches of physics, interesting in their own
right\,\cite{Kubo1991,Tuckerman2010}, 
and their development 
should prove useful in various research areas, 
since these branches 
are closely related to nanotechnology\,\cite{Tasaki2001},
mathematical engineering including
Markov chain Monte Carlo methods\cite{Andrieu2003}, 
and so on\,\cite{Kita2010,Goto2020}. Nonequilibrium phenomena 
are time-dependent thermodynamic phenomena, and some simple cases
have successfully been addressed\,\cite{Zubarev1,Zubarev2}. 
Although intricate systems are never fully appreciated,
some progress in understanding such has been made in proper frameworks.  
An example is on a dynamical process from a metastable
state into a most stable equilibrium state\,\cite{Campa2009}.
Another example is to deal with hysteresis curves
in magnetic systems\,\cite{Mayergoyz1986}.  
Beyond these, one should expect further progress.   
As stated above,
although considerable activity is being devoted to formulate nonequilibrium 
statistical thermodynamics, there is little consensus in the literature on
its foundation. To establish a foundation of nonequilibrium theory,
one might focus on a canonical example as a first step. This is because 
the Ising model, a canonical model,
played a central role in developing equilibrium theory\,\cite{Toda1983}.
Choosing a simple toy model appropriately 
many quantities are evaluated analytically, and insights can be gained from its
simplicity.  
One of such a good model is the Husimi-Temperley model.
This is based on
the Ising model, by modifying the interaction range 
from the nearest neighbor one to the 
global or mean field type. 
This model, the Husimi-Temperley model, 
exhibits a phase transition, and several 
quantities can be calculated
analytically\,\cite{Brankov1983,Mori2010,Matsueda2014}.  
On the one hand,  
one might be concerned that such long-range interactions 
are physically irrelevant. On the other hand, 
one might think that 
systems with long-range interactions are ubiquitous in the 
world. Examples of such systems  
include self-gravitating particles, two-dimensional fluid, and so 
on\,\cite{Campa2009}.
Since these examples exist, the study of a class of statistical systems with
long-range interactions is meaningful, where this class includes spin systems
with long-range interactions. %
In this paper this later perspective is 
adapted. 

To formulate nonequilibrium statistical mechanics one might
take the attitude that a contact geometric approach is employed\,
\cite{Grmela2014,Goto2015JMP,Haslach1997}. 
One of the questions in constructing such a formulation 
is how to deal with phase transitions\,\cite{Quevedo2011}, 
and another one is how to introduce dynamics describing nonequilibrium
phenomena. 
In this paper both of these questions are addressed for the case of
the Husimi-Temperley model, that is an Ising type model.  
This model enables many quantities to be expressed analytically,
and because of its simplicity physical insights can be gained.
One advantage of 
the use of contact geometry is that 
Legendre singularity theory\,\cite{Arnold1990} 
is expected to provide a sophisticated tool set to elucidate mechanisms of
phase transitions\,\cite{Aicardi2001,Aicardi2002}.  

\subsection*{Outline of this contribution}
In this paper analysis of the Husimi-Temperley model is
summarized to make this paper self-contained, and then 
its geometric description is proposed. In this proposal 
stable and unstable segments of the hysteresis and pseudo-free energy curves
are considered.  
Each curve is   
identified with a union of 
pruned segments of a 
projected Legendre submanifold,    
where this Legendre submanifold is a $1$-dimensional 
submanifold of a $3$-dimensional  
contact manifold. 
In this framework a thermodynamic phase space is identified with
  a contact manifold where the contact form restricts
  vector fields so that the first law of thermodynamics holds, and
  the time-development of contact Hamiltonian systems is identified with
  the time-development of thermodynamic processes
  in thermodynamic phase spaces.  
The main theorem in this paper 
  and its physical interpretation are 
  informally stated as follows. 
\begin{Claim}
  \label{fact-claim-main}
  (informal version of Theorem\,\ref{claim-cubic-contact-Hamiltonian3}). 
  The integral curves of a contact Hamiltonian vector field
  on a $3$-dimensional 
  contact manifold connect a unstable segment of a 
  Legendre submanifold and a stable one. 
Physically this contact Hamiltonian vector field expresses  
the dynamical process departing from
a unstable branch of the hysteresis curve to a stable one.
Simultaneously this vector field expresses the dynamical process
departing from a unstable branch of the free-energy curve to a stable one. 

\end{Claim}
 
To explain how to arrive at this claim, a procedure together with
calculations of this paper is summarized below. Since this 
summary is an abstraction of the calculations for the specific model, 
this summary is seen as a generalization from the specific procedure.
\begin{enumerate}
  \item
    Introduce a statistical model (of spins) with
    a parameter $J_{\,0}\in\mbbR$,  
    let $x\,\in\mbbR$ and $y\,\in\mbbR$ 
denote a dimensionless applied external field and 
magnetization, respectively. Then    
    $x$ and $y$ form a pair of thermodynamic conjugate variables.
    In general, the main task for elucidating thermodynamic
      properties of a microscopic model is
      to calculate the corresponding partition function by integrating all 
      the degrees of freedom with some measure. This measure is often chosen
      as the canonical measure,
      and the partition function yields the free-energy.
      Consider the case that 
    a dimensionless free-energy per degree of freedom $\psi_{\,J_0}$ 
is obtained with the so-called saddle point method:
$$
\psi_{\,J_0}(x,y^{\,*})
\simeq \min_{y}\ \psi_{\,J_0}(x,y)
=\min_{\mu}\ \psi_{\,J_0}(x,y_{\,\mu}^{\,*}),
$$
where $\mu\,\in\mbbN$ is a label for
discriminating various local minima of $\psi_{\,J_0}$ with respect to $y$, and
$y_{\,\mu}^{\,*}$ denotes a local minimum point.  
In this paper the following form of  $\psi_{\,J_0}$ is focused:  
$$
\psi_{\,J_0}(x,y)
=J_{\,0}y^{\,2}-\int \check{s}(\Delta)\,\dr\,\Delta,\qquad
\Delta
:=2J_{\,0}y+x,
$$
where $\check{s}$ is a function, and
$\check{s}(\Delta)=\tanh(\Delta)$  for the Husimi-Temperley model.  
\item
  Introduce the $3$-dimensional contact manifold
  $(T^{\,*}\mbbR\times\mbbR,\lambda)$
  whose coordinates are
$(x,y,z)$ so that $\lambda=\dr z+y\,\dr x$ is a contact $1$-form. 
Then a union of (metastable) equilibrium states are identified with a 
Legendre submanifold. The coordinate expression of
a (metastable) equilibrium state is    
$(x,y_{\mu}^{\,*}(x),z(x,y_{\,\mu}^{\,*}(x)))$ labeled by $\mu$, where 
$$
\frac{\partial \psi_{\,J_0}}{\partial y_{\,\mu}^{\,*}}
=2\left(y_{\,\mu}^{\,*}-\check{s}(\Delta_{\,\mu}^{\,*})\,\right) J_{\,0}
=0,\quad
\frac{\partial\psi_{\,J_0}}{\partial x}
=-y_{\,\mu}^{\,*},\quad
z=\psi_{\,J_0}(x,y_{\,\mu}^{\,*}),\qquad
\Delta_{\,\mu}^{\,*}
:=2J_{\,0}y_{\,\mu}^{\,*}+x,
$$
 with $\check{s}$ constituting a self-consistent equation
$y_{\,\mu}^{\,*}=\check{s}(\Delta_{\,\mu}^{\,*})$. 
Self-consistent equations are often derived  
in the study of 
systems with phase transitions, 
       where a phase transition is equivalent to 
       a bifurcation of the solution of the self-consistent equation.
From this construction, the hysteresis curve is 
nothing but the projection of the Legendre submanifold onto
the $(x,y)$-plane up to sign convention. 
In addition, the pseudo-free energy curve is the projection of
the Legendre submanifold onto the $(x,z)$-plane. 
The set of multiple branches of the hysteresis curve
is recognized as a multi-valued function of $x$,
$x\mapsto y_{\,\mu}^{\,*}=y_{\,\mu}^{\,*}(x)$. 
If this multi-valued function is invertible, 
then $x=x(y_{\,\mu}^{\,*})$ exists,
where the function $y_{\,\mu}^{\,*}\mapsto x(y_{\,\mu}^{\,*})$  
may be a single-valued function. 
In the Husimi-Temperley model
$x$ can explicitly be written in terms of 
$y_{\mu}^{\,*}$ on the Legendre submanifold,  
and the multiple branches of the hysteresis curve can be drawn by
varying the value of $y_{\,\mu}^{\,*}$ continuously
in the recognition that the graph
$(x,y_{\,\mu}^{\,*}(x))$ is depicted by
$(x(y_{\,\mu}^{\,*}),y_{\,\mu}^{\,*})$, where
$x:y_{\,\mu}^{\,*}\mapsto x(y_{\,\mu}^{\,*})$ is a single-valued function.
Thus the
Legendre submanifold whose projections are labeled by $\mu$ 
can be treated as a one submanifold, 
rather than multiple submanifolds.  
The projection of the Legendre submanifold onto the $(x,z)$-plane expresses
a pseudo-free energy as a multi-valued function. This
multi-valued function expresses a set
of metastable, unstable, and most stable equilibrium  states.
These projections in the symmetry broken phase
are depicted in Fig.\,\ref{picture-outline-unpruned-Legendre}. Note that
$\psi_{\,J_0}$ is not convex with respect to $x$\,
(see Remark\,\ref{remark-properties-of-psi} and Ref.\,\cite{Shimizu2007}),
whereas $\psi_{\,J_{0}}$ is convex with respect
to $y$ in the high temperature phase\,
(see Remark\,\ref{remark-properties-of-psi} together with 
Fig.\,\ref{Husimi-free-energy-fixed-x-picture}).  

\begin{figure}[htb]
\begin{picture}(120,65)
\unitlength 1mm
\put(41,20){$z$}
\put(61,2){$x$}
\put(54,15){$z=\psi_{\,J_0}(x(y_{\,\mu}^{\,*}),y_{\,\mu}^{\,*})$}
\put(39.5,4){\line(0,1){15}}
\put(20,4){\line(1,0){40}}
\put(54,5){\line(-2,1){30}}
\put(25,5){\line(2,1){30}}
\qbezier(24,20.2)(40,11)(56,20.6)
\put(109,21){$y$}
\put(131,10){$x$}
\put(131,17){$x=x(y_{\,\mu}^{\,*})$} 
\put(109,2){\line(0,1){17}}
\put(90,11){\line(1,0){40}}
\put(90,4){\line(1,0){10}}
\qbezier(100,4)(117,4)(118,8)
\qbezier(100,14)(101,18)(118,18)
\put(118,18){\line(1,0){10}}
\qbezier(100,14)(100.2,12)(109,11)
\qbezier(109,11)(118.8,10)(118,8)
%
%
\end{picture}
\caption{Unpruned projections of the Legendre submanifold 
  in the low temperature phase (symmetry broken phase). 
  (Left) The $(x,z)$-plane. 
  (Right) The $(x,y)$-plane.
  }
\label{picture-outline-unpruned-Legendre}
\end{figure}
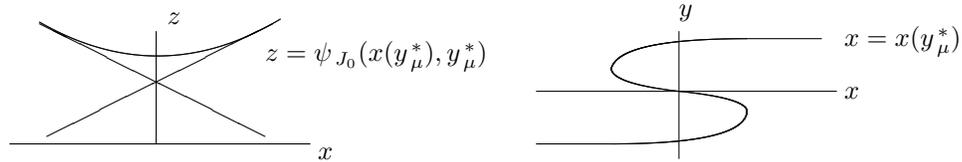

\item
Prune the top branch on the $(x,z)$-plane
of the projection 
(see Fig.\,\ref{picture-outline-pruned-Legendre}).
This pruning procedure is equivalent to prune the middle segment
passing through $(0,0)$ on the $(x,y)$-plane.
Then the resultant disconnected segments 
of the projection of the Legendre submanifold
yield disconnected hysteresis  and
pseudo-free energy curves that 
are  expected to be observed in experiments. 

\begin{figure}[htb]
\begin{picture}(120,65)
\unitlength 1mm
\put(41,20){$z$}
\put(61,2){$x$}
\put(57,20){$z=\psi_{\,2}(x)$}
\put(59,16){defined on $\cI^{\,+}$}
\put(50,10){$z=\psi_{\,1}(x)$}
\put(55,6){defined on $\cI^{\,+}$}
\put(4,20){$z=\psi_{\,2}(x)$}
\put(5,16){defined on $\cI^{\,-}$}
\put(4,10){$z=\psi_{\,1}(x)$}
\put(5,6){defined on $\cI^{\,-}$}
\put(39.5,4){\line(0,1){15}}
\put(20,4){\line(1,0){40}}
\put(30,0){$\cI^{\,-}$}%
\put(45,0){$\cI^{\,+}$}%
\put(54,5){\line(-2,1){30}}
\put(25,5){\line(2,1){30}}
\put(109,21){$y$}
\put(131,10){$x$}
\put(131,17){$x=x(y_{\,\mu}^{\,*})$}
\put(109,2){\line(0,1){17}}
\put(90,11){\line(1,0){40}}
\put(95,7){$\cI^{\,-}$}
\put(120.5,7){$\cI^{\,+}$}
\put(90,4){\line(1,0){10}}
\qbezier(100,4)(117,4)(118,8)
\qbezier(100,14)(101,18)(118,18)
\put(118,18){\line(1,0){10}}
\linethickness{0.5mm}
\put(25.0,3.5){\line(1,0){14}}
\put(40.0,3.5){\line(1,0){14}}
\put(100.5,10.5){\line(1,0){8}}
\put(109.5,10.5){\line(1,0){8}}
\end{picture}
\caption{Pruned segments of the projections of the Legendre submanifold 
  in the low temperature phase (symmetry broken phase). 
  The regions $\cI^{\,\mp}$ are such that
    $\cI^{\,-}:=\{\, x\in\mbbR_{\,<0}\,|\,\psi_{\,1}(x)<\psi_{\,2}(x)\,\} \subset\mbbR$ and
    $\cI^{\,+}:=\{\, x\in\mbbR_{\,>0}\,|\,\psi_{\,1}(x)<\psi_{\,2}(x)\,\} \subset\mbbR$,    
  where $\psi_{\,\mu}(x)=\psi_{\,J_0}(x,y_{\,\mu}^{\,*}(x))$
  is an abbreviation.    
  (Left) The $(x,z)$-plane. 
  (Right) The $(x,y)$-plane.
  }
\label{picture-outline-pruned-Legendre}
\end{figure}
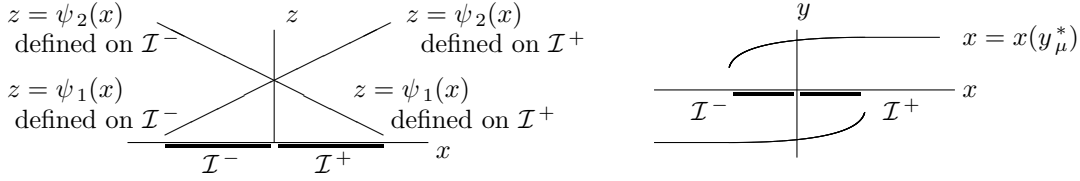

\item
  Choose a contact Hamiltonian
  $$
  h(x,y,z)
  =-\,\psi_{\,0}(x)(z-\psi_{\,1}(x))(z-\psi_{\,2}(x))^{\,2},
  $$
  where $\psi_{\,0}$ is a positive function of $x$. 
  Its corresponding   
  contact Hamiltonian vector field expresses the dynamical process 
  on the $(x,y)$- and $(x,z)$-planes. 
  The pruned projections of the Legendre submanifold 
  are shown to be stable and unstable 
  (see Fig.\,\ref{picture-outline-vector-field}).
  This gives Claim\,\ref{fact-claim-main}.

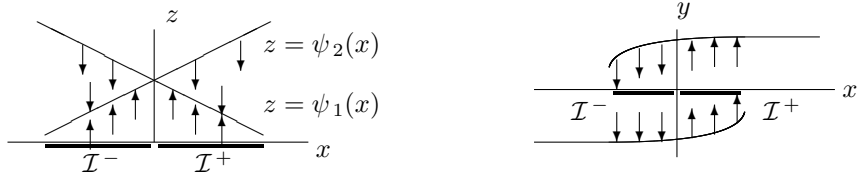
\begin{figure}[htb]
\begin{picture}(120,65)
\unitlength 1mm
\put(41,20){$z$}
\put(61,2){$x$}
\put(54,16){$z=\psi_{\,2}(x)$}
\put(54,8){$z=\psi_{\,1}(x)$}
\put(39.5,4){\line(0,1){15}}
\put(20,4){\line(1,0){40}}
\put(30,0){$\cI^{\,-}$}
\put(45,0){$\cI^{\,+}$}
\put(54,5){\line(-2,1){30}}
\put(25,5){\line(2,1){30}}
\put(31,3){\vector(0,1){4}}
\put(31,12){\vector(0,-1){4}}
\put(30,17){\vector(0,-1){4}}
\put(34,5){\vector(0,1){4}}
\put(34,15){\vector(0,-1){4}}
\put(37,7){\vector(0,1){4}}
\put(42,7){\vector(0,1){4}}
\put(45,5){\vector(0,1){4}}
\put(45,15){\vector(0,-1){4}}
\put(48.5,3.5){\vector(0,1){4}}
\put(48.5,11.5){\vector(0,-1){4}}
\put(51,17.5){\vector(0,-1){4}}
\put(109,21){$y$}
\put(131,10){$x$}
\put(109,2){\line(0,1){17}}
\put(90,11){\line(1,0){40}}
\put(95,7){$\cI^{\,-}$}
\put(120.5,7){$\cI^{\,+}$}
\put(90,4){\line(1,0){10}}
\qbezier(100,4)(117,4)(118,8)
\qbezier(100,14)(101,18)(118,18)
\put(118,18){\line(1,0){10}}
\put(101,8){\vector(0,-1){4}}
\put(101,15){\vector(0,-1){4}}
\put(104,8){\vector(0,-1){4}}
\put(104,16.5){\vector(0,-1){4}}
\put(107,8){\vector(0,-1){4}}
\put(107,17){\vector(0,-1){4}}
\put(111,4.5){\vector(0,1){4}}
\put(111,13.5){\vector(0,1){4}}
\put(114,5){\vector(0,1){4}}
\put(114,14){\vector(0,1){4}}
\put(117,6.5){\vector(0,1){4}}
\put(117,14){\vector(0,1){4}}
\linethickness{0.5mm}
\put(25.0,3.5){\line(1,0){14}}
\put(40.0,3.5){\line(1,0){14}}
\put(100.5,10.5){\line(1,0){8}}
\put(109.5,10.5){\line(1,0){8}}
\end{picture}
\caption{Contact Hamiltonian vector field that expresses the dynamical process
  in the low temperature phase (symmetry broken phase). 
  The fixed point sets 
  are pruned segments of the projections of the 
  Legendre submanifold.  
  (Left) The $(x,z)$-plane. 
  (Right) The $(x,y)$-plane.
  }
\label{picture-outline-vector-field}
\end{figure}

\end{enumerate}

In the case where there is only one single-valued function of $x$
one can find a contact Hamiltonian  
such that the corresponding segment of the projection of the
Legendre submanifold is 
attractor as has been argued in Refs.\,\cite{Goto2015JMP,Entov2021}.

As a corollary of Claim\,\ref{fact-claim-main},
the following is obtained.
\begin{Claim}
  \label{fact-claim-sub-a}
  (informal version of
  Corollary\,\ref{fact-combining-the-contact-Hamiltonian-system-h3}).   
  When pruning the unstable segments of the projected Legendre submanifold, 
  as the set of attractors of the contact Hamiltonian systems,  
  the cusp of the shape $\wedge$ appears on the $(x,z)$-plane,   
  and
  the kink of the shape $\step$ appears on the $(x,y)$-plane 
  (see Fig.\,\ref{picture-outline-cusp-kink}).
  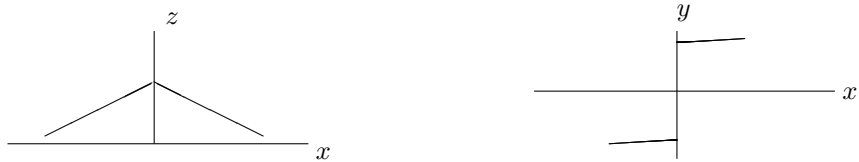
\begin{figure}[htb]
\begin{picture}(120,65)
\unitlength 1mm
\put(41,20){$z$}
\put(61,2){$x$}
\put(39.5,4){\line(0,1){15}}
\put(20,4){\line(1,0){40}}
\put(54,5){\line(-2,1){14.5}}
\put(25,5){\line(2,1){14.2}}
\put(109,21){$y$}
\put(131,10){$x$}
\put(109,2){\line(0,1){17}}
\put(90,11){\line(1,0){40}}
\qbezier(100,4)(108,4.5)(109,4.5)
\qbezier(109,17.5)(110,17.5)(118,18)
%
\end{picture}
\caption{Union of the stable pruned projections of the Legendre submanifold 
  in the low temperature phase (symmetry broken phase). 
  (Left) The $(x,z)$-plane. The cusp of the shape $\wedge$ appears, and 
  the existence of this cusp corresponds to the existence of the 
  $1$st-order phase transition.
  (Right) The $(x,y)$-plane. The kink of the shape $\step$ appears, and
  this expresses the experimental observation
  where the hysteresis curve is ruined by perturbation.   
  }
\label{picture-outline-cusp-kink}
\end{figure}
\end{Claim}
  From Claim\,\ref{fact-claim-sub-a}, the long-time evolution of the
  proposing contact Hamiltonian system plays a similar role of
  the Maxwell construction discussed in thermodynamics and the role of the 
  convexification by the Legendre transforms.

The rest of this paper is organized as follows.
In Section\,\ref{section-preliminaries}, some preliminaries are provided 
in order to keep this paper self-contained. They are 
basics of contact geometry 
including projections of Legendre submanifolds,
contact Hamiltonian systems, and
so on. In addition the so-called Husimi-Temperley model and
its thermodynamics are briefly summarized.  
In Section\,\ref{section-geometry-dynamical-process},
after arguing that  
metastable and unstable  
equilibrium states are described as a Legendre 
submanifold in a contact manifold, 
pruned segments of the projected Legendre submanifold are introduced.
Then, a contact Hamiltonian vector field is introduced where this vector field
expresses the dynamical process under the case that the 
unstable and stable segments of the hysteresis curve exist
in the symmetry broken phase.   
Section\,\ref{Conclusions} summarizes
this paper and  discusses some future works.
In Appendix\,\ref{section-appendix}, 
some other contact Hamiltonian systems are discussed.

\section{Preliminaries}
\label{section-preliminaries}

This section is intended to provide a  
brief summary of the necessary background, 
and consists of $2$ parts. 
They are about contact geometry, and about thermodynamic properties of the 
Husimi-Temperley model.

\subsection{Contact and symplectic geometries}
\label{section-preliminary-geometry}
To argue contact geometry of nonequilibrium thermodynamics,
some known facts on contact and symplectic geometries
are summarized and notations are fixed here\,\cite{Libermann1987,Silva2008}. 
Various formulas and 
tools developed in differential geometry are known\,\cite{Frenkel,Nakahara}.  
For example, the Lie derivative of a $k$-form $\alpha$ along a vector 
field $X$ can be written as
$\cL_{X}\alpha=\dr\ii_{\,X}\alpha+\ii_{\,X}\dr \alpha$, where 
$\dr$ is the exterior derivative and $\ii_{\,X}$ the interior product with
$X$. 
This is known as the Cartan formula.

Let $\cC$ be a $(2n+1)$-dimensional manifold ($n=1,2,\ldots$), 
and $\lambda$ a $1$-form on $\cC$ such that
$$
\lambda\wedge\underbrace{\dr \lambda\wedge\cdots\wedge\dr\lambda}_{n}
\neq 0,\quad \mbox{at any point on $\cC$}.
$$
Then this $\lambda$ is referred to as a {\it contact form}.  
Notice that another $1$-form $f\,\lambda$ with $f$ 
being a non-vanishing 
function is also a contact form if $\lambda$ is contact. 
If a $2n$-dimensional vector space $E\subset T_{\,p}\cC$ is written by
$E=\ker\lambda$ with $\ker\lambda=\{\ X\in T_{p}\cC\ |\  \lambda(X)=0\ \}$
around $p\in\cC$, 
then the pair $(\cC,\ker \lambda)$ is referred to as a $(2n+1)$-dimensional
{\it contact manifold (in the wider sense)}, where $\lambda(X)$ denotes the
pairing between $\lambda$ and $X$. 
According to the Darboux theorem, there exist coordinates $(x,y,z)$ 
such that
\beq
\lambda
=\dr z-\sum_{a=1}^{n}y_{\,a}\,\dr x^{\,a},\quad\mbox{or}\quad
\lambda
=\dr z+\sum_{a=1}^{n}y_{\,a}\,\dr x^{\,a},
\label{canonical-contact-form}
\eeq
where $x=(x^{\,1},\ldots,x^{\,n})$ and $y=(y_{\,1},\ldots,y_{\,n})$.
These coordinates are referred to as {\it canonical} or
{\it Darboux coordinates}.   
If 
there exists such a contact form globally over $\cC$, then
the pair $(\cC,\lambda)$ is referred to as a
{\it contact manifold (in the narrower sense)}.
In Section\,\ref{section-geometry-dynamical-process} of 
  this paper, this case, $\lambda$ is globally defined, is considered  
 and contact manifolds are written as $(\cC,\lambda)$.  
One typical contact manifold is 
given as $T^{\,*}\mbbR^{n}\times\mbbR$ with some $1$-form. 

On a contact manifold $(\cC,\lambda)$, there exists a vector field $R$ that
satisfies
$$
\ii_{\,R}\,\dr \lambda
=0,\qquad\mbox{and}\qquad
\ii_{\,R}\lambda
=1.
$$
This $R$ is referred to as the {\it Reeb vector field},
and is uniquely determined for a fixed $\lambda$.
This is written as  
$R=\partial/\partial z$ in the canonical coordinates such that $\lambda$ is
written as \fr{canonical-contact-form}. 

A {\it contact vector field} $X$
is a vector field on a contact manifold
$(\cC,\lambda)$ that preserves 
the contact structure $\ker\lambda$, $\cL_{\,X}\lambda=f\lambda$
with $f$ being a non-vanishing function. 
There is a way to specify a contact vector field with a function described
below. 
The contact Hamiltonian vector field $X_{\,h}$ associated with a function  $h$
is the uniquely determined vector field such that
\beq
\ii_{X_{\,h}}\lambda
=h,\qquad
\lambda\wedge\cL_{X_{\,h}}\lambda
=0,
\label{contact-Hamiltonian-vector}
\eeq
where the second equation reduces to 
$$
\ii_{\,X_{h}}\dr\lambda
=-\,(\dr h-(Rh)\lambda),
$$
which is shown by applying $\ii_{\,R}$ and the Cartan formula.  
The function $h$ is referred to as a {\it contact Hamiltonian}.
Note that there are some sign conventions on defining
contact Hamiltonian vector fields. 
The coordinate expression of $X_{\,h}$ is obtained as follows. 
Let $(x,y,z)$ be canonical coordinates such that
$\lambda=\dr z-\sum_{a=1}^ny_{\,a}\dr x^{\,a}$ in \fr{canonical-contact-form}. 
Then from \fr{contact-Hamiltonian-vector}, one has   
$$
X_{\,h}
=\sum_{a=1}^{n}\left(\dot{x}^{\,a}\frac{\partial}{\partial x^{\,a}}
+\dot{y}_{\,a}\frac{\partial}{\partial y_{\,a}}\right)
+\dot{z}\frac{\partial}{\partial z},
$$
where $\dot{x}$, $\dot{y}$, and $\dot{z}$ are the functions  
  $$
\dot{x}^{\,a}
=-\,\frac{\partial h}{\partial y_{\,a}},\quad
\dot{y}_{\,a}
=\frac{\partial h}{\partial x^{\,a}}+y_{\,a}\frac{\partial h}{\partial z},
\quad
\dot{z}
=h-\sum_{b=1}^{n}y_{\,b}\frac{\partial h}{\partial y_{\,b}},\qquad
a=1,\ldots,n.
$$
Identifying $\dot{\ }=\dr/\dr t$ and $t\in\cI$,
  $(\cI\subseteq\mbbR)$, 
  one has that 
  $X_{\,h}$ expresses a dynamical system.
  This $t$ will be identified with time in
    Section\,\ref{section-geometry-dynamical-process}. 
  This dynamical
  system is referred to as a {\it contact Hamiltonian system} associated
  with $h$.    
For $3$-dimensional contact manifolds, 
one can drop the subscripts and superscripts as 
$$
\dot{x}
=-\,\frac{\partial h}{\partial y},\quad
\dot{y}
=\frac{\partial h}{\partial x}+y\,\frac{\partial h}{\partial z},
\quad
\dot{z}
=h-y\,\frac{\partial h}{\partial y}.
$$
In the case of $\lambda=\dr z+y\,\dr x$ for $3$-dimensional manifolds, 
one derives 
\beq
\dot{x}
=\frac{\partial h}{\partial y},\quad
\dot{y}
=-\,\frac{\partial h}{\partial x}+y\,\frac{\partial h}{\partial z},
\quad
\dot{z}
=h-y\,\frac{\partial h}{\partial y}.
\label{contact-hamiltonian-vector-field-component}
\eeq

On contact manifolds, some special submanifolds play important roles.
Given a $(2n+1)$-dimensional contact manifold $(\cC,\lambda)$, an
$n$-dimensional submanifold such that $\phi^{\,*}\lambda=0$ is
referred to as a {\it Legendre submanifold}
(\,{\it Legendrian submanifold}\,), where $\phi^{\,*}$ is the pullback of 
an embedding $\phi$. 
Such a submanifold is generated by a function, and an example is shown
in Example\,\ref{Example-Legendre-submanifold-by-function}.
If the dimension of a Legendre submanifold is unity, then this submanifold
is referred to as a {\it Legendre curve}. 
\begin{Example}
\label{Example-Legendre-submanifold-by-function}
  Let $(\mbbR^{\,3},\lambda)$ be a
  $3$-dimensional contact manifold, $(x,y,z)$ its coordinates, and
  $\lambda=\dr z\mp y\,\dr x$.
In addition, let $\psi_{\,\R}$ be a function of $x$. Then   
$$
\phi\,\cA_{\,\psi_{\R}}
=\left\{\ (x,y,z)\in\mbbR^{\,3}\ \bigg|\ z=\psi_{\,\R}(x),\quad\mbox{and}\quad
y=\pm\frac{\dr \psi_{\,\R}}{\dr x}\ 
\right\}
$$
is a Legendre submanifold generated by $\psi_{\,\R}$
due to $\phi^{\,*}\lambda=0$ and $\dim(\phi\,\cA_{\psi_{\R}})=1$. 
\end{Example}
Another example being relevant to this paper is as follows. 
\begin{Example}
  \label{example-Legendre-submanifold-consistent-equation}
  Let $(\mbbR^{\,3},\lambda)$ be a $3$-dimensional contact manifold, $(x,y,z)$
  its coordinates,
  and $\lambda=\dr z- y\,\dr x$. In addition,  
let $\psi_{\,\I}$, $f_{\,\I}$, and $\Delta$ be the functions 
$$
\psi_{\,\I}(x,y)
=y^{\,2}-f_{\,\I}(\Delta),\quad
\Delta(x,y)
=2 y- x.
$$
Then the embedded manifold 
\beq
\phi\,\cA_{\,\psi_{\,\I}}
=\left\{\ (x,y,z)\in\mbbR^{\,3}\ \bigg|\
z=\psi_{\,\I}(x,y),\quad\mbox{and}\quad
y=\frac{\dr f_{\,\I}}{\dr \Delta}(\Delta(x,y)),\quad
\mbox{where}\ x\in\cI 
\ \right\},
\label{Legendre-submanifold-f-Delta-psi}
\eeq
is a Legendre submanifold, where 
$y=\dr f_{\,\I}/\dr \Delta$ 
can be treated as an algebraic equation for $y$
with $x$ being a continuous parameter,
and $\cI\subset \mbbR$ is a region in which the
real solution for $y$ exists. 
The submanifold \fr{Legendre-submanifold-f-Delta-psi} is verified to
be Legendrian as $\dim(\phi\,\cA_{\,\psi_{\,\I}})=1$ and 
\beqa
\phi^{\,*}\lambda
&=&\dr\, (\phi^{\,*}z) - (\phi^{\,*}y)\,\dr (\phi^{\,*}x)
\non\\
&=&\frac{\partial \psi_{\,\I}}{\partial x}\dr x
+\frac{\partial \psi_{\,\I}}{\partial y}\dr y
-\frac{\dr f_{\,\I}}{\dr \Delta}\dr x
\non\\
&=&-\frac{\dr f_{\,\I}}{\dr\Delta}\frac{\partial\Delta}{\partial x}\dr x
+\left(2y-\frac{\dr f_{\,\I}}{\dr\Delta}\frac{\partial\Delta}{\partial y}
\right)\dr y- \frac{\dr f_{\,\I}}{\dr \Delta}\dr x
\non\\
&=&0.
\non
\eeqa
Similarly, for the case that
$$
\lambda=\dr z+y\,\dr x,\quad
\psi_{\,\I}=y^{\,2}-f_{\,\I}(\Delta),\quad
\Delta=-2\,y-x,
$$ 
the embedded submanifold
$$
\phi\,\cA_{\,\psi_{\,\I}}
=\left\{\ (x,y,z)\in\mbbR^{\,3}\ \bigg|\
z=\psi_{\,\I}(x,y),\quad\mbox{and}\quad
y=- \frac{\dr f_{\,\I}}{\dr \Delta}(\Delta(x,y)),\quad
\mbox{where}\ x\in\cI 
\ \right\},
$$
is Legendrian. 
In some physical context, the algebraic equation above is derived 
as a self-consistent equation for determining the value of an order parameter in
statistical mechanics
(see around \fr{Husimi-algebraic-equation-beta-x-y-mu-decompose}).
\end{Example}
In Example\,\ref{example-Legendre-submanifold-consistent-equation} 
with $\lambda=\dr z-y\,\dr x$,  
it follows that 
$$
\frac{\partial \psi_{\,\I}}{\partial y}
=2y-2\frac{\dr f_{\,\I}}{\dr\Delta}
=0\qquad
\mbox{on $\phi\cA_{\,\psi_{\I}}$ },
$$
implying that $y$ on the Legendre submanifold 
is written by the solution to $\partial \psi_{\,\I}/\partial y=0$, and that
the solution is written in terms of the derivative 
$y= \dr f_{\,\I}/\dr \Delta$. 
This structure motivates the following generalization
from Example\,\ref{example-Legendre-submanifold-consistent-equation}. 
\begin{Example}
\label{example-Legendre-submanifold-consistent-equation-psi-2}  
Let $(T^{\,*}\mbbR\times\mbbR,\lambda)$
  be a $3$-dimensional contact manifold with $\lambda=\dr z+y\,\dr x$ where
  $(x,y,z)$ its coordinates. In addition,  
  let $\psi_{\,\II}$ be a function of $(x,y)$. 
Then the embedded manifold expressed in coordinates 
\beq
\phi\,\cA_{\,\psi_{\,\II}}
=\left\{\ (x,y,z) 
\ \bigg|\
z=\psi_{\,\II}(x,y),\ 
y=-\frac{\partial \psi_{\,\II}}{\partial x},\
\frac{\partial\psi_{\,\II}}{\partial y}=0.
\quad \mbox{where}\ x\in\cI 
\ \right\},
\label{Legendre-submanifold-f-Delta-psi-2}
\eeq
is a Legendre submanifold, where
$\partial\psi_{\,\II}/\partial y=0$ can be 
 treated as an algebraic equation for $y$
with $x$ being a parameter, and $\cI\subset \mbbR$ is a region in which the
real solution for $y$ exists.
This submanifold is Legendrian due to $\dim(\phi\,\cA_{\,\psi_{\,\II}})=1$
and $\phi^{\,*}\lambda=0$.
\end{Example}

In what follows some projections of Legendre submanifolds 
are defined. 
Consider first the cotangent bundle $T^{\,*}\mbbR$.
Let $x$ be a coordinate of $\mbbR$, and
$y$ a coordinate of $T_{\,x}^{\,*}\mbbR$.
The so-called {\it Liouville $1$-form} is expressed as  
$\alpha=y\,\dr x$, inducing 
 a {\it symplectic form} expressed as   
$\omega=\dr \alpha=\dr y\wedge\dr x$. 
Second, let $z$ be a coordinate of another $\mbbR$. Then   
take the $3$-dimensional contact manifold, 
$(T^{\,*}\mbbR\times\mbbR,\lambda)$ where $\lambda=\dr z\mp\alpha$. 
On this contact manifold  
a Legendre submanifold, or Legendre curve,
is identified with  an embedded $1$-dimensional curve 
in the $3$-dimensional manifold, and its projection onto a plane could yield 
some singularities. The projection of the Legendre curve onto the $(x,y)$-plane is
referred to as a {\it Lagrange map}, and that onto the $(x,z)$-plane as a 
{\it Legendre map}. The image of a Legendre map is referred to as the
{\it wave front} of the Legendre curve. 
\begin{Example}
  For the case of $\lambda=\dr z\mp y\,\dr x$
  on $T^{\,*}\mbbR\times\mbbR$,  
the wave front of the Legendre curve
generated by $\psi_{\,\R}$ being a (single-valued) function depending
on $x$ 
$$
\phi\cA_{\,\psi_{\R}}
=\left\{\ (x,y,z)  
\ \bigg|\ y=\pm\frac{\dr \psi_{\,\R}}{\dr x}(x),\quad\mbox{and}\quad z=\psi_{\,\R}(x)
\  \right\}
$$
is the graph $(x,\psi_{\,\R}(x))$ on the $(x,z)$-plane. Here
there is no singularity
associated with this projection provided that $\psi_{\,\R}$ is 
smooth.    
\end{Example}
\begin{Example}
  For the case of $\lambda=\dr z-y\,\dr x$, the wave front of the Legendre curve
expressed in coordinates 
  \beq
  \phi\cA_{\,\psi_{\,\I}^{\,\prime}}
  =\left\{\ (x,y,z) 
  \ \bigg|\  
  z=y^2-\frac{\Delta^{\,3}}{3},\ y=\Delta^{\,2},\quad
  \mbox{where}\ 
\Delta(x,y)=2y-x,\ \mbox{and}\  \frac{-1}{8}\leq x\ 
\right\}
\label{picture-toy-example-Legendre-submanifold}
\eeq
is obtained as follows
(see Example\,\ref{example-Legendre-submanifold-consistent-equation}, 
and take $f(\Delta)=\Delta^{\,3}/3$ with $\lambda=\dr z-y\,\dr x$).  
First the conditions $y=\Delta^{\,2}=(2y-x)^{\,2}$, and $-1/8\leq x$ yield
on the $(x,y)$-plane as the image of the Lagrange curve  
$$
y_{\,\pm}(x)
=\frac{4x+1\pm \sqrt{1+8x}}{8},\quad\mbox{for}\quad
-\frac{1}{8} \leq x,
$$
which can be seen as a multi-valued function of $x$. Each branch of this
multi-valued function is jointed at $x=-1/8$, as can be verified from
$y_{\,+}(-1/8)=y_{\,-}(-1/8)=1/16$. 
Second,
$z=y^{\,2}-\Delta^{\,3}/3$ is drawn on the $(x,z)$-plane as 
$$
z(x,y_{\,\pm}(x))
=y_{\,\pm}(x)^{\,2}-\frac{y_{\,\pm}(x)^{\,3/2}}{3}.
$$
Hence, the wave front is 
$(x,z_{\,+}(x))\cup(x,z_{\,-}(x))$, where
$z_{\,\pm}(x)=z(x,y_{\,\pm}(x))$. 

In Fig.\ref{picture-example-wave-front}, the projections of the Legendre curve
onto various planes are shown.
A singular point appears at $x=-1/8$ on the wave front. 
The projection onto
the $(x,z)$-plane can be seen as a double-valued function. 
This view will be used in analyzing a dynamical process in
Section\,\ref{section-geometry-dynamical-process}. 

\begin{figure}[htb]
\centering
\includegraphics[width=5.2cm]{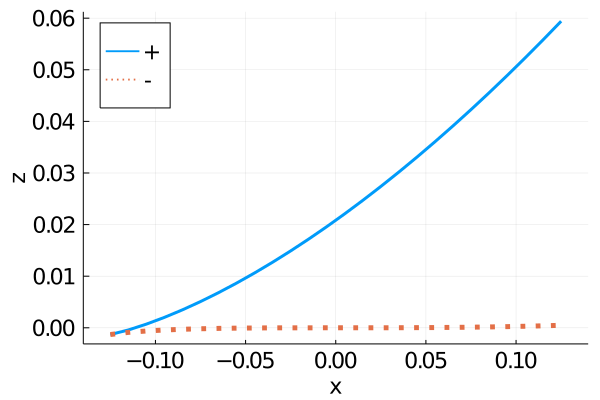}
\includegraphics[width=5.2cm]{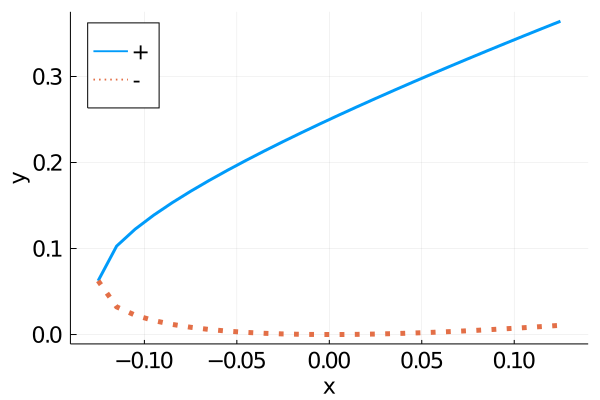}
\includegraphics[width=5.2cm]{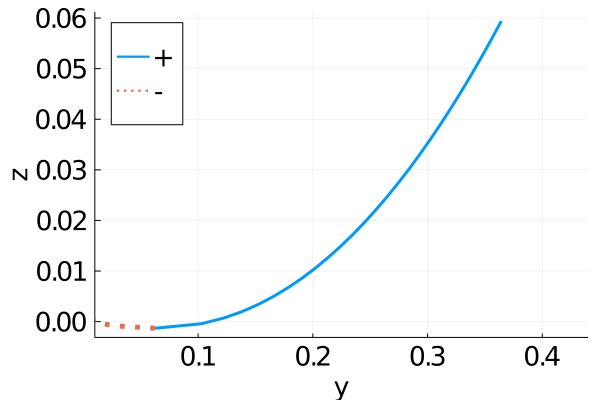}
\caption{
  Projection of \fr{picture-toy-example-Legendre-submanifold}, 
  where $+$ and $-$ denote the lines obtained by $y_{\,+}(x)$
  and $y_{\,-}(x)$, respectively.
  (Left) Projection onto the $(x,z)$-plane, the wave front.  
  (Middle) Projection onto the $(x,y)$-plane, the image of the Lagrange map.
  (Right) Projection onto the $(y,z)$-plane.
}
\label{picture-example-wave-front}
\end{figure}

\end{Example}

\subsection{Thermodynamics of the Husimi-Temperley model}
\label{section-thermodynamics-Ising}
In this subsection an Ising type spin system is introduced, and then 
its thermodynamic properties  derived with
canonical statistical mechanics
are 
summarized. The aim of this subsection is to introduce a toy model, 
where that model should be appropriate in the sense
that most of quantities are analytically obtained and
introduced quantities are physically interpretable.   
In Section\,\ref{section-geometry-dynamical-process}, 
geometric analysis of this model will be shown. 

Consider a lattice whose total number of lattice points is $N$. 
At the lattice point specified by $i\in\{1,\ldots,N\}\subset \mbbN$ put
a spin variable $\sigma_{\,i}=\pm 1$, and then 
$\sigma:=(\sigma_{\,1},\ldots,\sigma_{\,N})$.  
The space of the spin variables is denoted $\cS=\{\pm1\}^{\,N}$,
so that $\sigma\in\cS$.   
The total energy defined for this system is introduced as
\beq
\cH(\sigma)
=-\frac{J_{\,0}}{N}\sum_{i=1}^{N}\sum_{j=1}^{N}\sigma_{\,i}\sigma_{\,j}
-H\sum_{i=1}^{N}\sigma_{\,i},
\label{Husimi-Temperley-Hamiltonian}
\eeq
where $J_{\,0}\in\mbbR$ is constant expressing the 
strength of spin interactions, and 
$H\in\mbbR$ constant expressing an 
externally applied   
magnetic field. This $\cH(\sigma)$ has the physical
  dimension of energy by fixing the physical dimensions of $J_{\,0}$ and $H$.  
Equation\,\fr{Husimi-Temperley-Hamiltonian}
is seen as a function, $\cH:\cS\to\mbbR$.
This model is referred to as the
{\it Husimi-Temperley model}.  
To elucidate thermodynamic properties of the model, introduce $m:\cS\to\mbbR$
such that
$$
m(\sigma)
=\frac{1}{N}\sum_{i=1}^N\sigma_{\,i}, 
$$
which is an order parameter. The variables $m$ and $H$ form a thermodynamic
conjugate pair.  

The canonical statistical mechanics is then applied to
the Husimi-Temperley model so that thermodynamic properties of this model
are elucidated, where the heat bath 
temperature is denoted by $T>0$.   
The main task is to calculate the partition function
$$
Z=\sum_{\sigma_{1}=\pm1}\cdots\sum_{\sigma_{N}=\pm1}
\e^{\ -\,\beta\cH(\sigma)}
=\sum_{\sigma_{1}=\pm1}\cdots\sum_{\sigma_{N}=\pm1}\ 
\e^{\ \beta J_{\,0}\left(\sum_{l=1}^{N}\sigma_{l}\right)^{\,2}/\,N}
\,\e^{\,\beta H\sum_{l=1}^{N}\sigma_{\,l}},
$$
where $\beta$ has been defined by $\beta=1/(k_{\B}T)$ with $k_{\,\B}$ being
the Boltzmann constant.  In the following the so-called 
{\it saddle point method} for Gaussian integral is applied so that
an approximate expression of $Z$ is obtained for $N\gg 1$. 
Under this approximation 
the free energy obtained from that $Z$ yields various thermodynamic
quantities by differentiation. 

The expression of $Z$ reduces as follows. The 
identity
$$
\e^{\ bs^{\,2} }
=\int_{\mbbR}\frac{\dr \varpi}{\sqrt{2\pi}}
\e^{\ -\varpi^{\ 2}+2\varpi\sqrt{b}\,s},\quad s\in\mbbR
$$
with substitutions $b=\beta J_{\,0}/N$ and $s=\sum_{a=1}^{N}\sigma_{a}$ 
yields 
$$
Z=\int_{\mbbR}\frac{\dr \varpi}{\sqrt{2\pi}}
\e^{\ -\varpi^{\ 2}}\left[
  2\cosh \left(2\varpi\sqrt{\frac{\beta J_{\,0}}{N}}+\beta H\right)
  \right]^{\,N}.
$$
Furthermore, the change of variables from $\varpi$ to $\xi$ so that 
$\varpi=\sqrt{N}\xi$, one has
$$
Z=\sqrt{\frac{N}{2\pi}}
\int_{\mbbR}\,\dr \xi
\exp\left[\,-\,N\left[\,
    \xi^{\,2}-\ln \left(2\,\cosh\left(2\xi\sqrt{\beta J_{\,0}}+\beta H\,
    \right)\,\right)
    \,\right]\right].
$$
In the limit $N\gg 1$, the expression of $Z$ above reduces to
$$
Z\simeq
\exp\left[\,-\,N\min_{\xi\in\mbbR}\left[\,
    \xi^{\,2}-\ln \left(2\,\cosh\left(
    2\xi\sqrt{\beta J_{\,0}}+\beta H\,\right)\,\right)\,\right]\right],
$$
where some irrelevant constants have been omitted. This approximation is
known as the saddle point method.
Under this approximation the free energy $\cF=-k_{\,\B}T\ln Z$
is 
$$
\cF\simeq
\cF^{\,\saddle},\quad
\cF^{\,\saddle}(T,H;\beta,J_{\,0})
=k_{\,\B}T N\min_{\xi\in\mbbR}\left[\,
  \xi^{\,2}-\ln \left(2\,\cosh\left(2\xi\sqrt{\beta J_{\,0}}+\beta H\,
  \right)\,\right)\,\right].
$$
This expression is rewritten by introducing the variable $y$ satisfying 
$\xi=y\sqrt{\beta J_{\,0}}$
as 
\beqa
\cF^{\,\saddle}(T,H;\beta,J_{\,0})
&=&k_{\,\B}T N\min_{y\in\mbbR}\left[\,
  \beta J_{\,0}y^{\,2}-\ln \left(2\,\cosh\left(2 \beta J_{\,0}y+
  \beta H\,
  \right)\,\right)\,\right]
\label{Husimi-F-saddle}
\\
&=&N
f_{\,\beta,J_{\,0}}(x,y^{\,*}),
\label{Husimi-F-saddle-divided-by-N}
\eeqa
where
\beqa
x&=&H,
\label{x=H}\\
y^{\,*}
&=&\arg\min_{y\in\mbbR} f_{\,\beta,J_{\,0}}(x,y),
\label{Husimi-y-saddle}\\
f_{\,\beta,J_{\,0}}(x,y)
&=&J_{\,0}y^{\,2}
-\frac{1}{\beta }\ln
\left(2\,\cosh\left(2\beta J_{\,0}\,y+\beta x\,\right)\,\right),
\label{f_beta-Husimi}
\eeqa
From \fr{Husimi-F-saddle-divided-by-N}, 
the physical meaning of $f_{\,\beta,J_{\,0}}(x,y^{\,*})$ is 
the value of the free energy per degree of freedom.   
From \fr{Husimi-F-saddle-divided-by-N} and \fr{f_beta-Husimi},
$f_{\,\beta,J_{\,0}}(x,y)$ 
can be interpreted as a relaxation or extension of
$f_{\,\beta,J_{\,0}}(x,y^{\,*})$, where $y^{\,*}$ is relaxed to $y\in\mbbR$.
This function $f_{\,\beta,J_{\,0}}:\mbbR^{\,2}\to\mbbR$ is referred
to as a {\it pseudo-free energy (per degree of freedom)} in this paper.
The dissimilarity between pseudo-free energy and free energy is that
the convexity of pseudo-free energy is not guaranteed.   

\begin{figure}[ht]
\centering
\includegraphics[width=7.0cm]{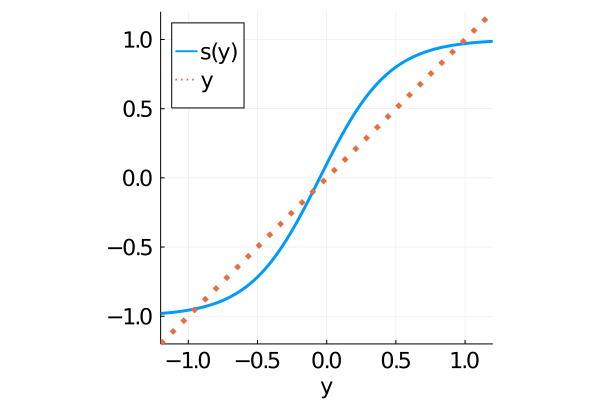}
\includegraphics[width=7.0cm]{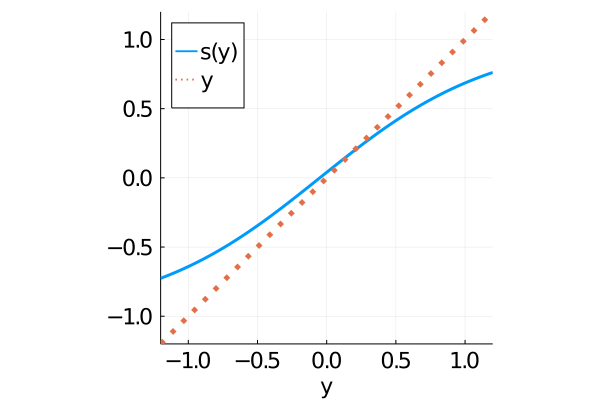}
\caption{
Intersection points of \fr{Husimi-algebraic-equation-beta-x-y-mu-decompose} 
are the solutions to  
\fr{Husimi-algebraic-equation-beta-x-y-mu}. 
(Left) There are $3$ intersection points in the low temperature phase
($\beta=1.0$, $J_{\,0}=1.0$, $x=0.1$), which
are denoted by $y_{\,1}^{\,*}, y_{\,2}^{\,*}, y_{\,3}^{\,*}$. 
(Right) There is $1$ intersection point in the high temperature phase
($\beta=0.4$, $J_{\,0}=1.0$, $x=0.1$), which is denoted by $y_{\,1}^{\,*}$. }
\label{Husimi-algebraic-equation-beta-x-y-mu-picture}
\end{figure}

The  $y^{\, *}$ obtained as \fr{Husimi-y-saddle} is chosen from
$\{y_{\,\mu}^{\,*}\,|\,\mu=1,2,3\}$,  
where $y_{\,\mu}^{\,*}$ satisfies
\beq
\left.\frac{\partial f_{\,\beta,J_{\,0}}}{\partial y}\right|_{y_{\,\mu}^{\,*}}
=0, 
\label{Husimi-algebraic-equation-beta-x-y-mu-0}
\eeq
with $\mu\in\mbbN$ denoting a label for
a solution to \fr{Husimi-algebraic-equation-beta-x-y-mu-0}.
The reason for introducing labels
$\mu$ is to 
take into account the possibility that the algebraic equation
\fr{Husimi-algebraic-equation-beta-x-y-mu-0} 
has at most countably many solutions.
This equation, \fr{Husimi-algebraic-equation-beta-x-y-mu-0}, is explicitly
expressed as 
\beq
y_{\,\mu}^{\,*}-\tanh(2\beta J_{\,0}y_{\,\mu}^{\,*}+\beta x\,)
=0,\qquad \mu=1,2,\ldots.
\label{Husimi-algebraic-equation-beta-x-y-mu}
\eeq
Notice that $y^{\,*}$ satisfies
\fr{Husimi-algebraic-equation-beta-x-y-mu}, since $y^{\,*}$ is chosen from
possible $y_{\,\mu}^{\,*}$, $(\mu=1,2,\ldots)$.  
One way to solve \fr{Husimi-algebraic-equation-beta-x-y-mu} is to
find intersection points of the curve and the line on the $(y,s)$-plane, 
\beq
s_{\,x;\beta,J_{\,0}}(y)
=\tanh(2\beta J_{\,0}y+\beta x\,),\qquad
\mbox{and}\qquad
s(y)
=y.
\label{Husimi-algebraic-equation-beta-x-y-mu-decompose}
\eeq  
In Fig.\,\ref{Husimi-algebraic-equation-beta-x-y-mu-picture},
the intersection points discussed above are shown.  

\begin{figure}[ht]
\centering
\includegraphics[width=7.0cm]{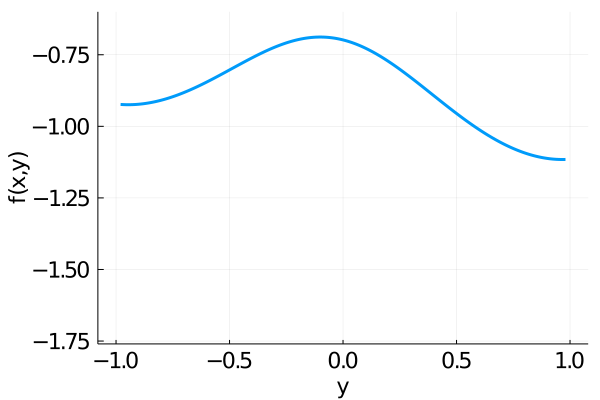}
\includegraphics[width=7.0cm]{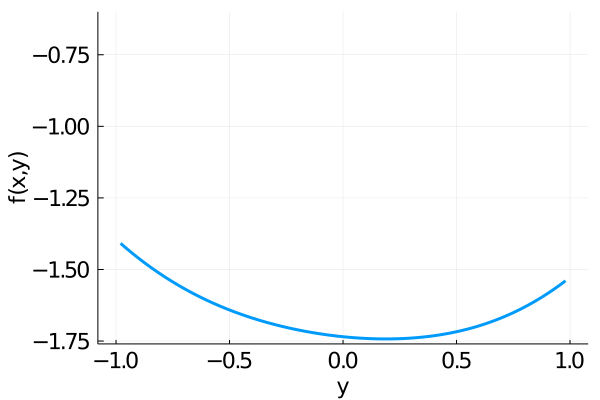}
\caption{Graph of $f_{\,\beta,J_{\,0}}(x,y)$ as a function of $y$ with
  $x$ kept fixed. The point $y^{\,*}(x)$ as a particular point of the $y$-axis 
  assigns the equilibrium 
  thermodynamic state for each $x$, where  
  $y^{\,*}(x)=\arg\min_{\mu}f_{\,\beta,J_{\,0}}(x,y_{\,\mu}^{\,*})$ with 
  $y_{\,\mu}^{\,*}$ being solutions to $\partial f_{\,\beta,J_{\,0}}/\partial y=0$,
  (see \fr{Husimi-algebraic-equation-beta-x-y-mu-0}). 
(Left) 
There are $3$ solutions $y_{\,\mu}^{\,*}$  in the low temperature phase
($\beta=1.0$, $J_{\,0}=1.0$, $x=0.1$), which
are denoted by $y_{\,1}^{\,*}, y_{\,2}^{\,*}, y_{\,3}^{\,*}$. 
(Right) There is $1$ solution $y_{\,\mu}^{\,*}$ in the high temperature phase
($\beta=0.4$, $J_{\,0}=1.0$, $x=0.1$), which is denoted by $y_{\,1}^{\,*}$.}
\label{Husimi-free-energy-fixed-x-picture}
\end{figure}

Although an explicit expression for $y_{\,\mu}^{\,*}$ as a function of
$x,\beta,J_{\,0}$ is not obtained,   
a condition when the number of solutions changes is found, that is obtained
from  
\beq
\left.\frac{\dr s_{\,x;\beta,J_{\,0}}}{\dr y}\right|_{\,y=\wt{y}^{\,*}}
\quad
\left\{
\begin{array}{cl}
>1&\mbox{the number of solutions is $3$,}\\
=1&\mbox{the critical point,}\\
<1&\mbox{the number of solutions is $1$ }
\end{array}
\right.,
\label{Husimi-critical-point-condition}
\eeq
where $\wt{y}^{\,*}$ is the solution of 
\fr{Husimi-algebraic-equation-beta-x-y-mu} near $y=0$. 
Hence the critical point is determined by the tangency condition
$\dr s_{\,x;\beta,J_{\,0}}/\dr y=1$ at $y=\wt{y}^{\,*}$,  
and is expressed as  
\beq
\left.\frac{\dr s_{\,x;\beta,J_{\,0}}}{\dr y}\right|_{\,y=\wt{y}^{\,*}}
=\left.\frac{2\beta J_{\,0}}{\cosh^{\,2}(2\beta J_{\,0}y+\beta x)}
\right|_{\,y=\wt{y}^{\,*}}
=1.
\label{Husimi-critical-point-condition-explicit}
\eeq
For example, consider the systems without external magnetic field,
$x=0$, 
from which $\wt{y}^{\,*}=0$.  This and
\fr{Husimi-critical-point-condition-explicit} yield that $\beta=1/(2J_{\,0})$ 
is the critical point, at which the number of the solutions changes. 
This critical point 
associated with a phase transition with respect to temperature  
divides $1$-dimensional  
region $\mbbR_{>0}$ into 
two, where this domain is the totality of $\beta$. 
One is the {\it low temperature phase}, and the other one
the {\it high temperature phase}.
The low temperature phase is also referred to as 
the {\it ordered phase} and as {\it symmetry broken  phase}.  
For each phase,  $f_{\,\beta,J_{\,0}}(x,y)$
as a function of $y$ is  
shown in Fig.\ref{Husimi-free-energy-fixed-x-picture}. 
The solution found around $y=0$ of $s_{\,x;\beta,J_{\,0}}(y)=s(y)$ in  
\fr{Husimi-algebraic-equation-beta-x-y-mu-decompose} is approximately
expressed as
$$
y\sim\frac{\beta x}{1-2\beta J_{\,0}},\quad \mbox{around $y=0$}, 
$$
where the Taylor expansion of $\tanh(\cdot)$ 
around the origin has been used.  

The variable $y^{\,*}$ is interpreted as the
canonical average of $m$, denoted 
$\avg{m}_{\can}$,
$$
  \avg{m}_{\can}
  =\frac{1}{Z}
  \sum_{\sigma_{\,1}=\pm1}\cdots\sum_{\sigma_{\,N}=\pm1} 
  m(\sigma)\,\e^{\,-\,\beta\cH(\sigma)}.
  $$
  To verify this interpretation, one first expresses   
$\avg{m}_{\can}$ in terms of a derivative of $\cF$. Then, the resultant
expression is shown to be written in terms of $y^{\,*}$. First, it follows that 
$$
\avg{m}_{\can}
=\frac{1}{N}\avg{\sum_{i=1}^{N}\sigma_{\,i}}_{\can}
=\frac{1}{N}\frac{\partial \ln Z}{\partial (\beta H)}
=\frac{-1}{N}\frac{\partial (\beta \cF)}{\partial (\beta H)}.
$$
Under the approximation
$$
\beta\cF\simeq\beta\cF^{\,\saddle}
=\beta N f_{\,\beta,J_{\,0}}(x,y^{\,*}),
$$
together with \fr{f_beta-Husimi}
and \fr{Husimi-algebraic-equation-beta-x-y-mu}, 
one has the desired relation, 
$$
\avg{m}_{\can}
\simeq
-\,\frac{\partial (\beta f_{\,\beta,J_{\,0}}(x,y^{\,*}))}{\partial (\beta x)}
=\tanh(2\beta J_{\,0}y^{\,*}+\beta x)
=\,y^{\,*}.
$$
Since $\avg{m}_{\can}\simeq y^{\,*}$ and 
$$
-1\leq \avg{m}_{\can}\leq 1,
$$
the $y^{\,*}$ is physically interpretable if 
$$ 
y^{\,*}\in \ol{\Upsilon},\quad
\Upsilon
=(-1,1),
$$
where $\ol{\Upsilon}=[-1,1]$.  
To discuss properties of the pseudo-free energy, 
restrict ourselves to the case that  
$\{\, y_{\,\mu}^{\,*}\,|\,\mu=1,2,3\, \}\subset \Upsilon$ and  
that $y_{\,\mu}^{\,*}\in\Upsilon$ depends on $x$ smoothly for each $\mu$.   
In this case, from \fr{Husimi-algebraic-equation-beta-x-y-mu} and the relation 
$$
\tanh^{\,-1}(\varsigma)
=\frac{1}{2}\ln\frac{1+\varsigma}{1-\varsigma}, \qquad
-1<\varsigma<1,
$$
that this $x$ can be written as a function of $y_{\,\mu}^{\,*}$,
$x:\Upsilon\to\mbbR$ such that  
\beq
x(y_{\,\mu}^{\,*})
=-2J_{\,0} y_{\,\mu}^{\,*} + \frac{1}{2\beta}
\ln\left(
\frac{1+y_{\,\mu}^{\,*}}{1-y_{\,\mu}^{\,*}}
\right),\qquad y_{\,\mu}^{\,*}\in\Upsilon.
\label{Husimi-algebraic-equation-beta-x-y-mu-inverse}
\eeq
The curve $y_{\,\mu}^{\,*} \mapsto x(y_{\,\mu}^{\,*})$  
is a single-valued function of $y_{\,\mu}^{\,*}$, and has the property that
$x(-y_{\,\mu}^{\,*})=-x(y_{\,\mu}^{\,*})$.  
Around $y_{\,\mu}^{\,*}=0$, 
this curve is approximately expressed as the line 
$$
x(y_{\,\mu}^{\,*})
\sim
\frac{1-2 \beta J_{\,0}}{\beta} y_{\,\mu}^{\,*},\qquad
\quad\mbox{around $y_{\,\mu}^{\,*}=0$}, 
$$
where the Taylor expansion of $\ln(1+\cdot)$ has been used.  

To discuss various quantities without any physical dimension, one 
introduces
\beq
\psi_{\,\ol{J_{0}}}(\ol{x},\ol{y})
:=\ol{J_{\,0}} \ol{y}^{\,2} -\ln(2\cosh(2\ol{J_{\,0}}\ol{y}+\ol{x})),\qquad
\ol{J_{\,0}}
:=\beta J_{\,0},\quad
\ol{x}
:=\beta x,\quad
\ol{y}(\ol{x})
:= y(\beta x),
\label{Husimi-psi-dimensionless}
\eeq
from \fr{x=H}, \fr{Husimi-y-saddle}, and \fr{f_beta-Husimi}.
Note that 
$\psi_{\,\ol{J_{0}}}(\ol{x},\ol{y})$ can be written as  
$\beta f_{\,\ol{J_{0}}}(\ol{x},\ol{y})$, and that 
$\ol{y}=y$ due to the property that $y$ is  dimensionless.
Similarly, $\ol{y_{\,\mu}^{\,*}}(\ol{x}):= y_{\,\mu}^{\,*}(\beta x)$,
and from the definition of $\ol{y_{\,\mu}^{\,*}}$ and
\fr{Husimi-algebraic-equation-beta-x-y-mu} it follows that 
\beq
\ol{y_{\,\mu}^{\,*}}
-\tanh(2\ol{J_{\,0}}\ol{y_{\,\mu}^{\,*}} +\ol{x})
=0,
\label{Husimi-algebraic-equation-beta-x-y-mu-dimensionless}
\eeq
which can be written as  
\beq 
\ol{x}(\ol{y_{\,\mu}^{\,*}})
=-\, 2\ol{J_{\,0}} \ol{y_{\,\mu}^{\,*}} + \frac{1}{2}
\ln\left(
\frac{1+\ol{y_{\,\mu}^{\,*}}}{1-\ol{y_{\,\mu}^{\,*}}}
\right),\qquad \ol{y_{\,\mu}^{\,*}}\in\Upsilon.
\label{Husimi-algebraic-equation-beta-x-y-mu-inverse-dimensionless}
\eeq
The  graph $(\ol{x},\ol{y_{\,\mu}^{\,*}}(\ol{x}))$ can be depicted
with \fr{Husimi-algebraic-equation-beta-x-y-mu-inverse-dimensionless}, and
this curve passes from 
$(-\infty,-1)$, via $(0,0)$, to $(+\infty,+1)$ on the $(\ol{x},\ol{y})$-plane. 
It is convenient to introduce the function 
$$
s_{\,\ol{x};\ol{J_{\,0}}}(\ol{y})
=\tanh (2\ol{J_{\,0}}\ol{y}+\ol{x}),
$$
that corresponds to $s_{\,x;\beta,J_{\,0}}(y)$ in 
\fr{Husimi-algebraic-equation-beta-x-y-mu-decompose}.

Differentiation of the above equations yields 
the following: 
\beqa
-\frac{\partial \psi_{\,\ol{J_{0}}}(\ol{x},\ol{y_{\,\mu}^{\,*}})}{\partial \ol{x}}
&=&\tanh(2\ol{J_{\,0}}\ol{y_{\,\mu}^{\,*}}+\ol{x})
=\ol{y_{\,\mu}^{\,*}},
\label{Husimi-psi-x-derivative}\\
\frac{\partial \psi_{\,\ol{J_{0}}}(\ol{x},\ol{y})}{\partial \ol{y}}
&=&2\ol{J_{\,0}}\left(\,
\ol{y}-\tanh(2\ol{J_{\,0}}\ol{y}+\ol{x})\,
\right),
\label{Husimi-psi-y-derivative}\\
\frac{\dr\ol{x}}{\dr \ol{y_{\,\mu}^{\,*}}}
&=&-\left(2\ol{J_{\,0}} - \frac{1}{1-\ol{y_{\,\mu}^{\,*}}^{\,2}}\right).
\label{Husimi-x-y-derivative}
\eeqa

\begin{Remark}
\label{remark-properties-of-psi}
  Observe from 
\fr{Husimi-critical-point-condition}--\fr{Husimi-critical-point-condition-explicit}
and 
\fr{Husimi-psi-x-derivative}--\fr{Husimi-x-y-derivative} 
the following.  
\begin{enumerate}
\item
For each $\ol{y_{\,\mu}^{\,*}}$, the function
$\psi_{\,\ol{J_{0}}}(\ol{x},\ol{y_{\,\mu}^{\,*}})$ is not convex with respect
to $\ol{x}$   
due to the derivative of \fr{Husimi-psi-x-derivative}, 
$$
\frac{\partial^2 \psi_{\,\ol{J_{0}}}(\ol{x},\ol{y_{\,\mu}^{\,*}})}{\partial \ol{x}^2}
=\frac{-1}{\cosh^{2}(2\ol{J_{\,0}}\ol{y_{\,\mu}^{\,*}}+\ol{x})}<0.
$$
\item
In the low temperature phase, 
the function
$\psi_{\,\ol{J_{0}}}(\ol{x},\ol{y})$ is not convex with respect to $\ol{y}$, 
due to the derivative of \fr{Husimi-psi-y-derivative}, 
$$
\frac{\partial^2 \psi_{\,\ol{J_{0}}}(\ol{x},\ol{y})}{\partial \ol{y}^2}
=2\ol{J_{\,0}}\left(\,
1-\frac{2\ol{J_{\,0}}}{\cosh^2(2\ol{J_{\,0}}\ol{y}+\ol{x})}\,
\right).
$$
\item
  When $2\ol{J_{\,0}}\geq 1$, it follows 
from $1/(1-\ol{y_{\,\mu}^{\,*}}^{\,2})\geq 1$ in \fr{Husimi-x-y-derivative}
that 
$\dr\ol{x}/\dr \ol{y_{\,\mu}^{\,*}}$ 
can vanish,
\beq
\frac{\dr\ol{x}}{\dr \ol{y_{\,\mu}^{\,*}}}
=0\quad\mbox{at}\quad
\ol{y_{\,\mu}^{\,*}}_{\,\pm},\quad
\ol{y_{\,\mu}^{\,*}}_{\,\pm}
:=\pm\sqrt{1-\frac{1}{2\ol{J_{\,0}}}}.
\label{spinodal-points-dimensionless}
\eeq
Hence in the region $2\ol{J_{\,0}}\geq 1$, 
the quantity $\dr \ol{y_{\,\mu}^{\,*}}/\dr\ol{x}$ can diverge, where
the $\dr \ol{y_{\,\mu}^{\,*}}/\dr\ol{x}$ is the negative of
the normalized magnetic susceptibility. 
\end{enumerate}

\end{Remark}

To avoid cumbersome notation we drop the bar, $\ol{\cdots }$, in what follows.
\begin{Remark}
Observe that the $y_{\,\mu}^{\,*}$ is a solution to 
the algebraic equation $\partial \psi_{\,J_{0}}/\partial y=0$, due to 
\fr{Husimi-psi-y-derivative}. This solution $y_{\,\mu}^{\,*}$ is written
as the negative of the derivative $-\partial \psi_{\,J_{0}}/\partial x$,  
due to  \fr{Husimi-psi-x-derivative}. 
This structure for $x$ and $y$ has also appeared in
Example\,\ref{example-Legendre-submanifold-consistent-equation-psi-2}.   
\end{Remark}
Physical interpretations of states
specified with $y^{\,*}$ and $y_{\,\mu}^{\,*}$ are assumed. 
 
\begin{Postulate}
\label{postulate-physical-interpretation-quasi-equilibrium}
(metastable and unstable equilibrium states). 
Fix $J_{\,0}(\neq 0)$ 
and $x$. When $y=y_{\,\mu}^{\,*}(x)$ and
$z=\psi_{\,J_{\,0}}(x,y_{\,\mu}^{\,*}(x))$
  with
  $y_{\,\mu}^{\,*}\neq y^{\,*}
  =\arg\min_{\mu^{\prime}}\psi_{\,J_{\,0}}(x,y_{\,\mu^{\prime}}^{\,*})$,
  the state specified by
  $(x,y_{\,\mu}^{\,*}(x),\psi_{\,J_{\,0}}(x,y_{\,\mu}^{\,*}(x)))$
  is assumed to express
  a metastable or unstable equilibrium state labeled with $\mu$. 
  In addition, when  $y=y^{\,*}$ and
  $z=\psi_{\,J_{\,0}}(x,y^{\,*}(x))$, the 
  state $(x,y^{\,*}(x),\psi_{\,J_{\,0}}(x,y^{\,*}(x)))$ 
  is assumed to express the (most-stable)
  equilibrium state.
\end{Postulate}
In Postulate\,\ref{postulate-physical-interpretation-quasi-equilibrium}, 
the terms ``metastable equilibrium state'' 
  and ``unstable equilibrium state'' have 
been written, and they are briefly
explained here. 
In this paper, equilibrium states are 
  special states where pairs of thermodynamic 
  variables can be described as the derivatives of a (multi-valued) potential.
By definition, there is a potential function defined at equilibrium states.
Equilibrium states are then classified with these potential
functions as follows. 
If the potential function is a single-valued function and convex,
then this function expresses the most stable equilibrium state.
If it is not the case, then such an equilibrium state is either
a metastable equilibrium state or a unstable equilibrium state. 
There is little consensus in the literature on how to define or to 
distinguish between metastable and unstable equilibrium states.
In this paper, 
the dissimilarity of the metastable and unstable equilibrium states is
that the unstable equilibrium states are not observed in experiments. 
In the conventional thermodynamics, the most stable equilibrium
state is constructed by the convexification with the Legendre transform. 

Definitions of metastable, unstable,  
and the most stable equilibrium states for the Husimi-Temperley model   
will be given in the language of contact geometry in the following section 
(see Definition\,\ref{definition-Husimi-equilibrium-Legendre-submanifold}).

From Postulate \ref{postulate-physical-interpretation-quasi-equilibrium},
the discussions in this subsection have been about
unstable and metastable equilibrium states
and the equilibrium state. 
So far no dynamical property of the system has been discussed. 
In the next section, dynamical equations will be proposed by giving
a contact Hamiltonian.

\section{Geometry of dynamical process in symmetry broken phase}
\label{section-geometry-dynamical-process}
In this section, a contact geometric description of
the thermodynamic variables derived from  
the Husimi-Temperley model
is given, and  
a physically appropriate dynamical system is proposed.  
To this end, physically allowed process are
discussed in terms of contact geometry first.

Consider a possible thermodynamic state specified by $(2n+1)$ 
variables, where the even number $2n$
is due to the pair of thermodynamic conjugate variables, and the $1$ due to 
the free energy value. 
During a change of thermodynamic states, the first law of thermodynamics
should hold. 
To discuss a smooth change of a state in time in terms of 
differential geometry, 
one introduces  a $(2n+1)$-dimensional manifold, a $1$-form, and 
a vector field on the manifold. This $1$-form
is used for restricting    
vector fields so that the first law of thermodynamics holds.  
From this discussion,
a contact manifold $(\cC,\lambda)$, or $(\cC,\ker\lambda)$ in a wider sense,  
and a class of vector fields are introduced for describing thermodynamics.
In this context,  $\cC=T^{\,*}Q\times\mbbR$ is a natural manifold with 
$Q$ being a manifold. 
On this setting 
an infinitesimal contact transform, that is a contact vector field,
gives physically allowed processes as curves by integrating the vector field.  
Thus, in this paper
\begin{itemize}
\item
a thermodynamic phase space is identified with a contact  manifold,   
\item
a thermodynamic state  is identified with a point of the manifold,
\item
  a dynamical thermodynamic process in a certain class of nonequilibrium
  processes is 
  identified with an integral curve of a contact vector field.
\end{itemize}
Beyond this formal procedure, for describing a particular
thermodynamic process or phenomenon, one specifies 
an appropriate contact vector field on the contact manifold. 
Choosing such an appropriate vector field from various allowed  
contact vector fields is not straightforward in general. 
Instead, rather than a vector field,  
one can alternatively choose a function,
since there is a correspondence between a function and a contact vector field,
where such a function is a contact Hamiltonian. 

For the Husimi-Temperley model, the thermodynamic 
phase space is specified as follows 
in this paper.
\begin{Definition}
\label{definition-Husimi-thermodynamic-space}
(thermodynamic phase space and contact manifold).  
Let $x$ be a coordinate for $\mbbR$, $y$ that for $T_{\,x}^{\,*}\mbbR$, and 
$z$ that for another $\mbbR$.
Take the $3$-dimensional manifold
$\cC=T^{\,*}\mbbR\times\mbbR$, and $\lambda$ the $1$-form  
$\lambda=\dr z+y\,\dr x$. This $\cC$ is 
referred to as the  thermodynamic phase space for the
Husimi-Temperley model. 
In addition the pair $(\cC,\lambda)$ is referred to
as the contact manifold for the Husimi-Temperley model. 
\end{Definition}
The  coordinates 
in Definition\,\ref{definition-Husimi-thermodynamic-space} at
the most stable  
equilibrium 
are interpreted as  
$x=\beta H$, $y\simeq \,\avg{m}_{\can}$, and $z$ is the lowest 
value of the dimensionless 
free-energy, $\beta f_{\,\beta,J_{\,0}}$, where $\simeq$ is due to the
saddle point approximation. Notice that entropy and temperature are not
included in Definition\,\ref{definition-Husimi-thermodynamic-space}, and
the magnetization and the externally applied magnetic field are focused in
this paper so that the dimension of the manifold is $3$, which renders various 
discussions on geometric properties simple. 
Temperature is then treated as a parameter, and thus
  all the curves in the thermodynamic phase
  space express isothermal processes in this paper. 

\subsection{Equilibrium}
Equilibrium states are the most fundamental states in thermodynamic
systems since they form the backbone of various thermodynamic states.
At equilibrium a thermodynamic 
quantity as a function
can be obtained by differentiating a potential with respect to
the corresponding thermodynamic conjugate variable.
In case of a gas system with 
constant temperature and volume environment, this potential is
the Helmholtz free energy. In case of systems of spins on lattices,
an appropriate potential is  
$\psi=\beta\cF$ with $\cF$ being 
the Gibbs free energy. 
From some arguments in thermodynamics, there is a set of 
correspondences between a fluid system contained in a box and a spin system. 
A magnetization and an applied external magnetic field 
in the spin system correspond to a volume and the negative of pressure
in the fluid system.

In the framework of contact geometric thermodynamics, 
an equilibrium state is described as a 
Legendre submanifold generated by a function, where such a function is
identified with a thermodynamic potential.  
For the Husimi-Temperley model, the
metastable, unstable, 
and the most stable equilibrium states are
defined as in a special case of 
Example\,\ref{example-Legendre-submanifold-consistent-equation-psi-2}.     
In the high temperature phase, the projection of the Legendre submanifold 
onto the $(x,z)$-plane can be expressed as 
a (single-valued) function, where the number of the 
labels is unity and thus the label can be omitted.   
Meanwhile in the symmetry broken phase, 
the projection of the Legendre submanifold 
onto the $(x,z)$-plane can be expressed as a $3$-valued function.
To discriminate these 
branches of this $3$-valued function, introduce the single-valued functions 
$\psi_{\,3}$, $\psi_{\,1}$, and $\psi_{\,2}$ such that 
$$
\psi_{\,J_{\,0}}(x,y)
=\left\{
\begin{array}{ll}
  \psi_{\,3}(x,y)&\mbox{the top branch on the $(x,z)$-plane}\\ 
  \psi_{\,1}(x,y)&\mbox{the bottom branch on the $(x,z)$-plane}\\
  \psi_{\,2}(x,y)&\mbox{the middle branch on the $(x,z)$-plane}
\end{array}
\right.\qquad\mbox{in the symmetry broken phase,}
$$
where the suffix $J_{\,0}$ for $\psi_{\,1}$, 
$\psi_{\,2}$, and $\psi_{\,3}$ has been omitted. 
\begin{Definition}
\label{definition-Husimi-equilibrium-Legendre-submanifold}
(metastable, unstable, and most stable equilibrium states).   
On the thermodynamic phase space $\cC$ of $(\cC,\lambda)$ in
Definition\,\ref{definition-Husimi-thermodynamic-space}, 
let $\psi_{\,J_{\,0}}$ be the function of $(x,y)$ as in  
\fr{Husimi-psi-dimensionless}. 
Then in the high temperature phase, the submanifold 
specified by $z=\psi_{\,J_{\,0}}$,
$y=y^{\,*}=-\,\partial \psi_{\,J_{\,0}}/\partial x$, and
$\partial \psi_{\,J_{\,0}}/\partial y|_{\,y=y^{\,*}}=0$ 
is the Legendre submanifold generated by $\psi_{\,J_{\,0}}$,
where $y^{\,*}$ 
is the unique solution 
that is written as the derivative of $\psi_{\,J_{\,0}}$, 
(see \fr{Husimi-psi-x-derivative}). 
This Legendre submanifold 
is referred to as the equilibrium state.  
In the symmetry broken phase,   
the Legendre submanifold   
with  $z=\psi_{\,1}$,  
$y=y_{\,1}^{\,*}=-\,\partial \psi_{\,1}/\partial x$, and 
$\partial \psi_{\,1}/\partial y|_{\,y=y_{\,1}^{\,*}}=0$  
is referred to as the {\it most stable equilibrium state}. 
The the Legendre submanifold with 
$z=\psi_{\,3}$,  
$y=y_{\,3}^{\,*}=-\,\partial \psi_{\,3}/\partial x$, and 
$\partial \psi_{\,3}/\partial y|_{\,y=y_{\,3}^{\,*}}=0$   
is referred to as the {\it unstable equilibrium state}. 
The Legendre submanifold with  
$z=\psi_{\,2}$,  
$y=y_{\,2}^{\,*}=-\,\partial \psi_{\,2}/\partial x$, and  
$\partial \psi_{\,2}/\partial y|_{\,y=y_{\,2}^{\,*}}=0$    
is referred to as the {\it metastable equilibrium state}. 

\end{Definition}
  Notice
  in Definition\,\ref{definition-Husimi-equilibrium-Legendre-submanifold}
  that, 
  although the number of the Legendre submanifold is $1$, there are $2$ 
  non-most-stable equilibrium states and $1$ the most stable equilibrium state
  for the Husimi-Temperley model.  
The metastable, unstable, and most stable equilibrium states
are originated from the 
Legendre submanifold, and are yielded by a classification and partition
of the submanifold. 

Several projections of Legendre submanifolds are defined in contact geometry
as have briefly been summarized
in Section\,\ref{section-preliminary-geometry}. 
In some cases singular points are described in a lower dimensional 
space and some multi-valued functions can be described. 
To detect phase transition and to 
characterize transitions in terms of 
contact geometry, 
such projections are applied to the equilibrium states of the
Husimi-Temperley model.
In the physics literature it is common to draw graphs on  
the $(y,z)$-plane. These graphs on the $(y,z)$-plane
correspond to the graphs in
Fig.\,\ref{Husimi-free-energy-fixed-x-picture} with some scaling factor.  
In the following other projections are focused. 

\begin{figure}[ht]
\centering
\includegraphics[width=5.3cm]{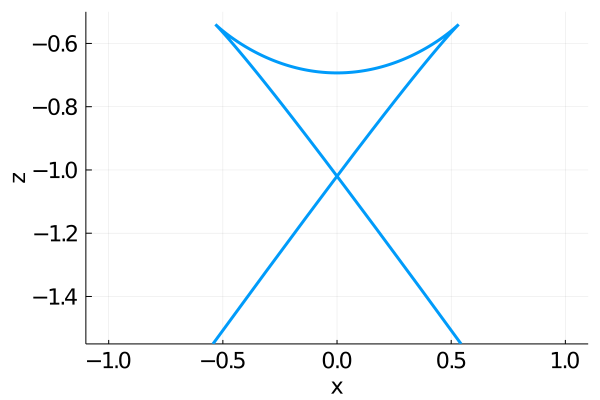}
\includegraphics[width=5.3cm]{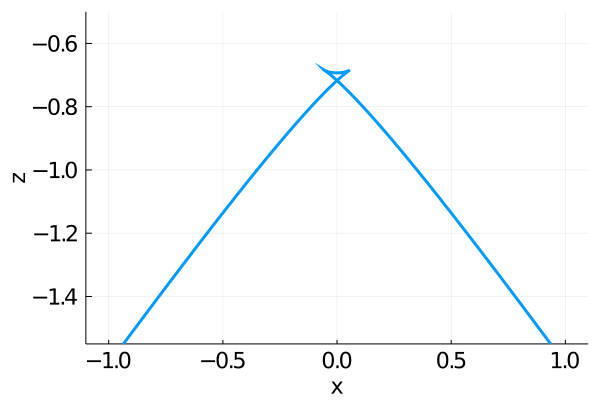}
\includegraphics[width=5.4cm]{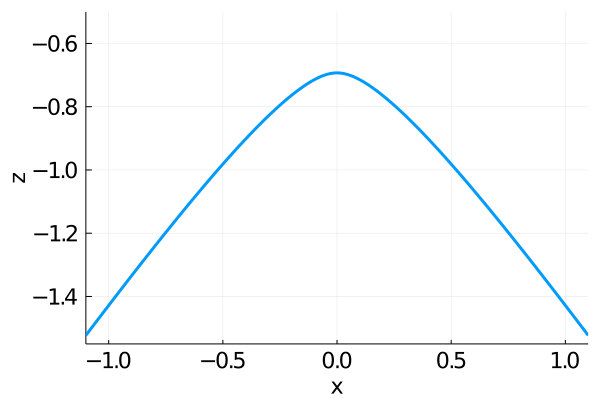}
\caption{Projections of the Legendre submanifold generated by
  $\psi_{\,J_{\,0}}(x,y)$ onto the $(x,z)$-plane (wave front).
  These were drawn with the use of 
  \fr{Husimi-algebraic-equation-beta-x-y-mu-inverse-dimensionless}
  and \fr{Husimi-psi-dimensionless}
  by varying the value of $y$ in $(-1,1)$. 
  (Left) Far from the critical point in the low temperature phase 
  ($J_{\,0}=1.0$ 
  in the dimensionless variable, obtained from 
  $\beta=1.0$ and $J_{\,0}=1.0$ as the dimensional variables), 
(Middle) Near the critical point in the low temperature phase  
  ($J_{\,0}=0.6$ %
  in the dimensionless variable, obtained from 
  $\beta=0.6$ and $J_{\,0}=1.0$ as the dimensional variables),  
  (Right) the high temperature phase
  ($J_{\,0}=0.4$  
  in the dimensionless variable, obtained from  
  $\beta=0.4$ and $J_{\,0}=1.0$ as the dimensional variables).}
\label{Husimi-wave-front-picture}
\end{figure}

In Fig.\,\ref{Husimi-wave-front-picture}, 
the $2$ cases of the wave front are depicted.
From this set of the cases, as 
known in the literature,  
the one 
in the lower temperature phase and the one in 
high temperature phase are distinguished. Such a difference is due to
a phase transition. 
In the framework of standard thermodynamics\,\cite{Callen},
the branch having the lowest value of the free energy is observed, 
and the ones having higher values are not observed. Hence 
the cusp of the wedge shape $\wedge$,
obtained by pruning the branches forming $\triangledown$  
in the left and middle panels, should appear in  
perturbed or noisy systems in experiments.
A physical interpretation of the 
branches having non-lowest values varies,
and ours is that those branches represent 
metastable and unstable equilibrium states\,(see Fig.2 of
  Ref.\cite{Aicardi2001}).  
In the case where the shape $\wedge$ appears,
the phase transition with respect to 
the externally applied field $H$   
is classified as the $1$st-order phase transition, since the free energy $z$
as a function 
is not differentiable at this cusp point. 

\begin{figure}[ht]
\centering
\includegraphics[width=5.3cm]{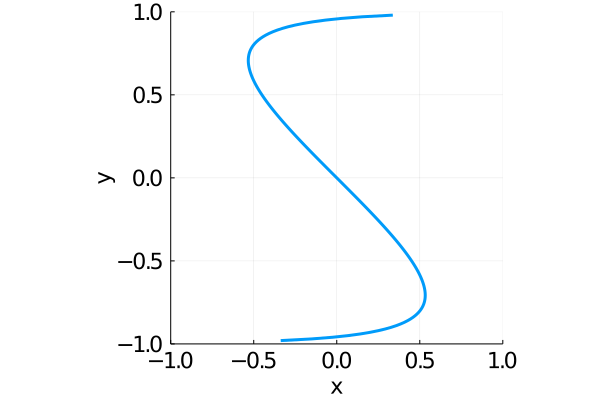}
\includegraphics[width=5.3cm]{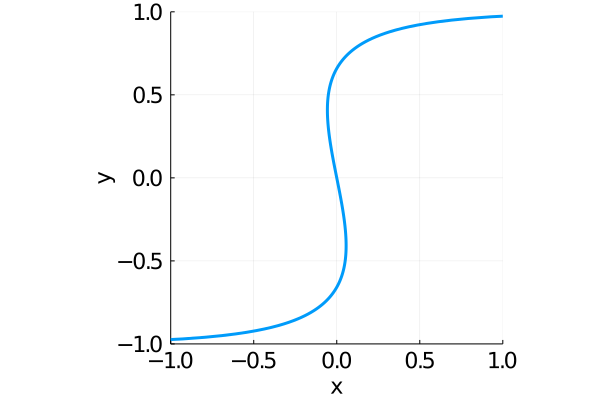}
\includegraphics[width=5.3cm]{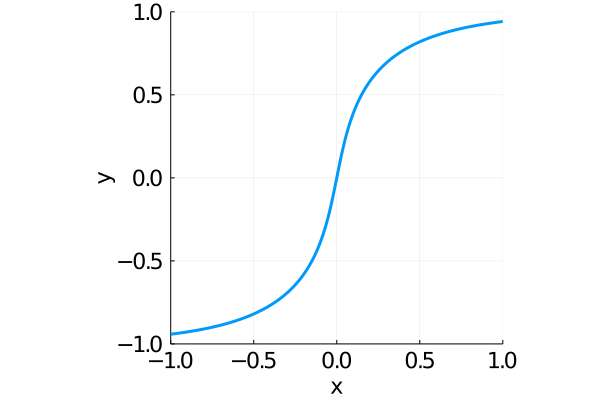}
\caption{Projections of the Legendre submanifold generated by
  $\psi_{\,J_{\,0}}(x,y)$  onto the $(x,y)$-plane (Images of the Lagrange map).
  These were drawn with the use of 
  \fr{Husimi-algebraic-equation-beta-x-y-mu-inverse-dimensionless} 
  by varying the value of $y$ in $(-1,1)$. 
  (Left) Far from the critical point in the low temperature phase 
  ($J_{\,0}=1.0$ 
  in the dimensionless variable constructed from 
  $\beta=1.0$ and $J_{\,0}=1.0$ in the dimensional variables), 
(Middle) Near the critical point in the low temperature phase 
  ($J_{\,0}=0.6$ %
  in the dimensionless variable, obtained from 
  $\beta=0.6$ and $J_{\,0}=1.0$ as the dimensional variables),  
(Right) The high temperature phase
  ($J_{\,0}=0.4$ 
  in the dimensionless variable, obtained from
  $\beta=0.4$ and $J_{\,0}=1.0$ as the dimensional variables).}
\label{Husimi-Legendre-map-picture}
\end{figure}

In Fig.\,\ref{Husimi-Legendre-map-picture}, the images of
the Lagrange map are drawn. As in the case of 
Fig.\,\ref{Husimi-wave-front-picture}, the lower temperature phase
differs from the higher temperature one, and forms    
a multi-valued function with the shape of $\mathcal{S}$. 
In  perturbed or noisy systems, such a multi-valued function
does not appear. 
One of observed structures in such experiments is
a kink structure of the shape $\step$. 
Another one, which we focus on first, is a pair of the 
disconnected curves that are obtained 
by pruning the middle segment passing through the origin $(0,0)$, since
such middle segment is physically unstable.
In this case, a hysteresis phenomenon takes place.
As will be discussed, from
Corollary\,\ref{fact-combining-the-contact-Hamiltonian-system-h3}, 
the kink $\step$ will be obtained as a stable fixed point set
in the contact manifold. 
 
\begin{figure}[htb]
\centering
\includegraphics[width=7.2cm]{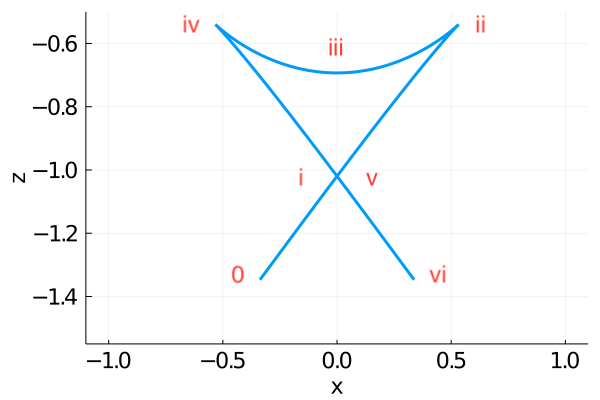}
\includegraphics[width=7.2cm]{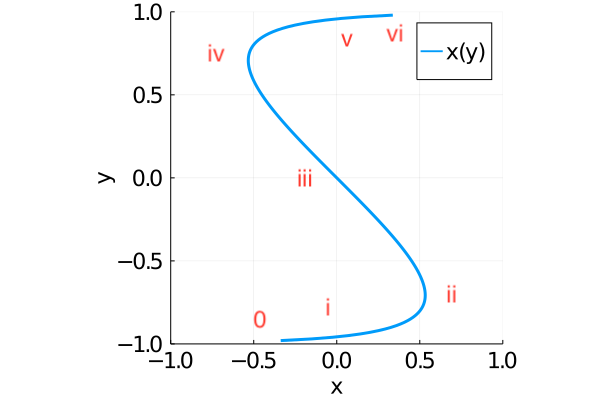}
\caption{Points of the projections shown in 
  Figs.\,\ref{Husimi-wave-front-picture} and 
  \ref{Husimi-Legendre-map-picture}.
The points $0,\ldots,\mathrm{vi}$ in the left figure correspond to  
the points $0,\ldots,\mathrm{vi}$ in the right one. 
(Left) The curve was drawn with the use of 
  \fr{Husimi-algebraic-equation-beta-x-y-mu-inverse-dimensionless}
  and \fr{Husimi-psi-dimensionless}
  by varying the value of $y$ in $(-1,1)$.
The undirected curves 
$\ol{\mathrm{v-vi}}$, $\ol{\mathrm{i-ii}}$, $\ol{\mathrm{ii-iii}}$ in the
left figure are 
the images of the functions $\psi_{\,1}$, $\psi_{\,2}$, and $\psi_{\,3}$, argued in
Section\,\ref{section-nonequilibrium} and Appendix\,\ref{section-appendix}.    
(Right) The line was drawn with the use of  
  \fr{Husimi-algebraic-equation-beta-x-y-mu-inverse-dimensionless} 
  by varying the value of $y$ in $(-1,1)$.
  The points $\mathrm{ii}$ and $\mathrm{iv}$ are   spinodal points, and they are expressed as 
  $(x(y_{\,\mu\,-}^{\,*}),y_{\,\mu\,-}^{\,*})$
  and $(x(y_{\,\mu\,+}^{\,*}),y_{\,\mu\,+}^{\,*})$. These spinodal points 
  $y_{\,\mu\,\pm}^{\,*}$ and $x$ as a function of $y$ have been defined in  
  \fr{spinodal-points-dimensionless}
  and \fr{Husimi-algebraic-equation-beta-x-y-mu-inverse-dimensionless},
  respectively.   
}
\label{Husimi-Legendre-hysteresis-picture}
\end{figure}

In Fig.\,\ref{Husimi-Legendre-hysteresis-picture},  
points of the projections shown in 
  Figs.\,\ref{Husimi-wave-front-picture} and 
  \ref{Husimi-Legendre-map-picture}
  are plotted for the low temperature phase.
  A {\it spinodal point} is the point where $\dr x/\dr y=0$ 
  in general,
  and in this case they are $\mathrm{ii}$ and
  $\mathrm{iv}$ on  the $(x,y)$-plane.
       In Section 9-4 of Ref.\cite{Callen} the Van der Waals gas system is considered, and
        a branch is identified with being ``unphysical''. 
       Then, in the Husimi-Temperley model,  
       the segment in between the spinodal points 
       is unphysical. In this paper unphysical states are assumed to be
       invisible or ruined. 
       In this sense the present projection of the curve does not reflect
       correct thermodynamics. 
       To render this segment non-existent, 
       introduce a pruned projection of
       the Legendre curve by removing such an invisible segment.
       The resultant pruned projection of
       the Legendre curve onto the $(x,y)$-plane consists of  
       disconnected curves. Similarly the resultant projection onto the
       $(x,z)$-plane consists of disconnected curves.      
 
\begin{figure}[htb]
\centering
\includegraphics[width=5.6cm]{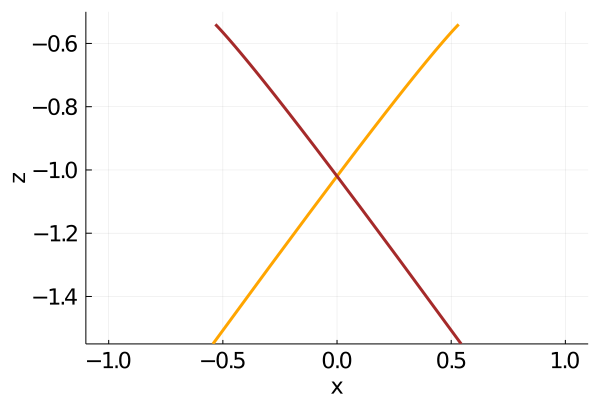}
\includegraphics[width=5.6cm]{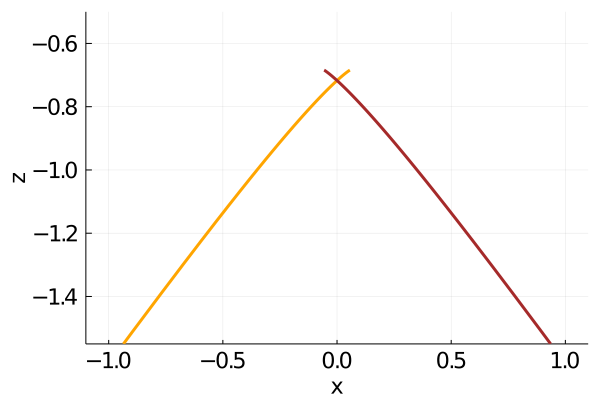}\\
\includegraphics[width=6.7cm]{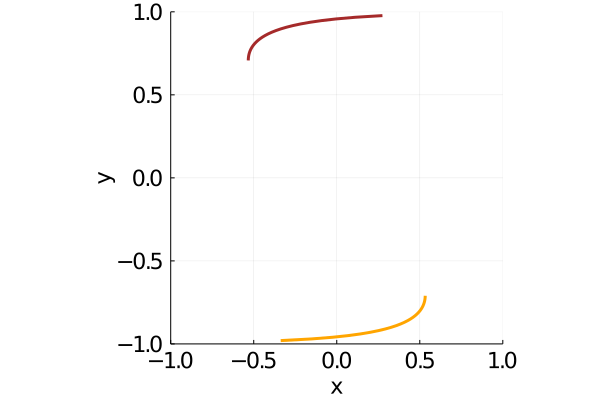}\hspace*{-16mm}
\includegraphics[width=6.7cm]{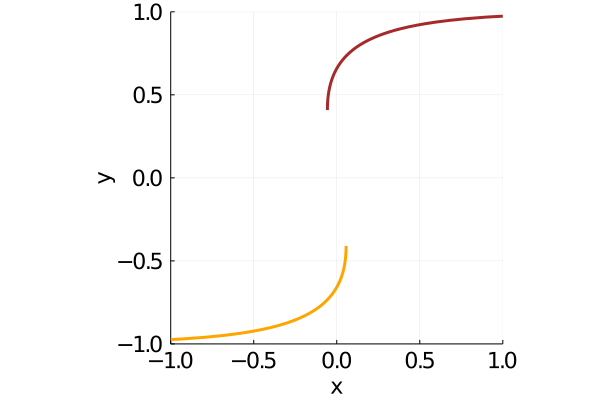}
\caption{Pruned projections of the Legendre submanifold generated by
  $\psi_{\,J_{\,0}}(x,y)$  onto the $(x,y)$- and $(x,z)$-planes.
  These were depicted with the use of 
  \fr{Husimi-algebraic-equation-beta-x-y-mu-inverse-dimensionless} 
  by varying the value of $y$ in $(-1,1)$, and pruning the middle segments. 
  Such middle parts on the $(x,y)$-plane 
  are from $(x(y_{\,\mu\,-}^{\,*}),y_{\,\mu\,-}^{\,*})$
  to $(x(y_{\,\mu\,+}^{\,*}),y_{\,\mu\,+}^{\,*})$, where
  $y_{\,\mu\,\pm}^{\,*}$ and $x$ as a function of $y$ have been defined in  
  \fr{spinodal-points-dimensionless}
  and \fr{Husimi-algebraic-equation-beta-x-y-mu-inverse-dimensionless},
  respectively. 
  The curves expressing lower values of $y$ in the lower panels of the left
  and right figures correspond to the curves of the shape of $/$
  in the upper panels. 
  (Left) Far from the critical point in the low temperature phase 
  ($J_{\,0}=1.0$ 
  in the dimensionless variable constructed from 
  $\beta=1.0$ and $J_{\,0}=1.0$ in the dimensional variables).
(Right) Near the critical point in the low temperature phase 
  ($J_{\,0}=0.6$ %
  in the dimensionless variable, obtained from 
  $\beta=0.6$ and $J_{\,0}=1.0$ as the dimensional variables).}  
\label{Husimi-partial-hysteresis1-picture}
\end{figure}

In Fig.\,\ref{Husimi-partial-hysteresis1-picture}
the pair of the disconnected curves is shown on the $(x,y)$-plane.
This pair of the curves is obtained
by pruning the middle segment passing through the origin $(x,y)=(0,0)$.
Edges of the pruned segments are located at the spinodal points 
$(x(y_{\,\mu\,-}^{\,*}),y_{\,\mu\,-}^{\,*})$ and
$(x(y_{\,\mu\,+}^{\,*}),y_{\,\mu\,+}^{\,*})$, where $y_{\,\mu\,\pm}^{\,*}$
have been defined in \fr{spinodal-points-dimensionless} and $x$
as a function of $y_{\,\mu}^{\,*}$ has been defined in
\fr{Husimi-algebraic-equation-beta-x-y-mu-inverse-dimensionless}. 
The corresponding pruned projections onto the $(x,z)$-planes form
double-valued functions.

  Singularities associated with the phase transition
  with respect to the external field 
  do not appear 
  in the $1$-dimensional Legendre submanifold $\phi\cA$
  embedded in the $3$-dimensional
  contact manifold. To verify this, first recall that in general  
a singular point of a curve is a point where its tangent vector vanishes.
Second, focus on the 
  curve as the Legendre submanifold $\gamma_{\,xyz}:\Upsilon\to\phi\cA$, 
  ($y\mapsto(x(y),y,z(y))$), where $y$ is the abbreviation of
  $y_{\,\mu}^{\,*}$ and $z(y)=\psi_{\,J_{0}}(x(y),y)$.  
  Then calculate the tangent vector along $\gamma_{\,xyz}$,
  from which one has  
\beqa
  \gamma_{\,xyz\,*}\left(\frac{\dr}{\dr y}\right)
  &=&\frac{\dr x}{\dr y}\frac{\partial }{\partial x}
    +\frac{\partial}{\partial y}
    +\frac{\dr z}{\dr y}\frac{\partial}{\partial z}\qquad
      =
\frac{\dr x}{\dr y}\frac{\partial }{\partial x}
    +\frac{\partial}{\partial y}
    +\left(\frac{\partial z}{\partial y}
    +\frac{\partial z}{\partial x}\frac{\dr x}{\dr y}
    \right)\frac{\partial}{\partial z}
\non\\
&=&\left(-2J_{\,0}+\frac{1}{1-y^{\,2}}\right)\frac{\partial}{\partial x}
+\frac{\partial}{\partial y}
  + \left(-2J_{\,0}+\frac{1}{1-y^{\,2}}\right)
  \frac{\partial\psi_{\,J_{0}}}{\partial x}\frac{\partial}{\partial z}
\neq 0,\quad \mbox{at any point of $y\in\Upsilon$,}  
\non
\eeqa
where \fr{Husimi-algebraic-equation-beta-x-y-mu-dimensionless},
\fr{Husimi-psi-y-derivative}, \fr{Husimi-x-y-derivative} have
been used, and 
$\gamma_{\,xyz\,*}$ is the push-forward of $\gamma_{\,xyz}$.
Thus, there is no singular point on the Legendre curve. 
Meanwhile such singularities appear on the $(x,z)$-plane
as the result of the projection. To show this, one   
calculates 
the tangent vector of the curve
$\gamma_{\,xz}:\Upsilon\to\mbbR\times\mbbR$, 
($y\mapsto (x(y),z(y))$),
\beqa
\gamma_{\,xz\,*}\left(\frac{\dr}{\dr y}\right)
&=&\frac{\dr x}{\dr y}\frac{\partial }{\partial x}
+\frac{\dr z}{\dr y}\frac{\partial}{\partial z}
\non\\
&=&\left(-2J_{\,0}+\frac{1}{1-y^{\,2}}\right)\frac{\partial}{\partial x}
  + \left(-2J_{\,0}+\frac{1}{1-y^{\,2}}\right)
  \frac{\partial\psi_{\,J_{0}}}{\partial x}\frac{\partial}{\partial z}.     
\non
\eeqa
From this calculation and \fr{spinodal-points-dimensionless}, one verifies
that there are singular points at the spinodal points, $y_{\,\mu\,\pm}^{\,*}$. 

\subsection{Nonequilibrium}
\label{section-nonequilibrium}
Nonequilibrium processes are time-dependent thermodynamic processes, 
and their geometric
descriptions have been proposed in the literature. 
 There are a variety of classes of nonequilibrium states, and
  our nonequilibrium thermodynamic states are such that 
  thermodynamic variables can uniquely specify thermodynamic states.  
In the contact geometric framework, 
such a description of a thermodynamic process 
is to choose a suitable contact Hamiltonian 
system. Among various nonequilibrium thermodynamic processes, 
relaxation processes have mainly been investigated,
where such a process describes a time-development of a state 
towards the most stable equilibrium state. 
In this section, an appropriate contact Hamiltonian is introduced for
describing 
the dynamical process from metastable equilibrium states
to the most stable equilibrium ones.

\begin{figure}[htb]
\begin{picture}(120,65)
\unitlength 1mm
\put(36,20){$z$}
\put(61,2){$x$}
\put(52,15){$z=\psi_{\,2}(x)$}
\put(52,8){$z=\psi_{\,1}(x)$}
\put(36,4){\line(0,1){15}}
\put(20,4){\line(1,0){40}}
\put(40,0){$\cI^{\,+}\ \subset\mbbR_{>0}$}
\put(58,5){\line(-4,1){34}}
\put(25,5){\line(2,1){30}}
\put(106,20){$z$}
\put(131,2){$x$}
\put(122,15){$z=\psi_{\,1}(x)$}
\put(108,4){\line(0,1){15}}
\put(90,4){\line(1,0){40}}
\qbezier(95,7)(108,25)(125,7)
\linethickness{0.5mm}
\put(36,3.5){\line(1,0){23}}

\end{picture}
\caption{Wave front. (Left) Low temperature phase.
  The label $\mu$ for $\psi_{\,\mu}$ is chosen so that
  $\psi_{\,1}(x)<\psi_{\,2}(x)$.  
  (Right) High temperature phase, the (single-valued) function appears.}
\label{Husimi-legndre-picture-labels2}
\end{figure}
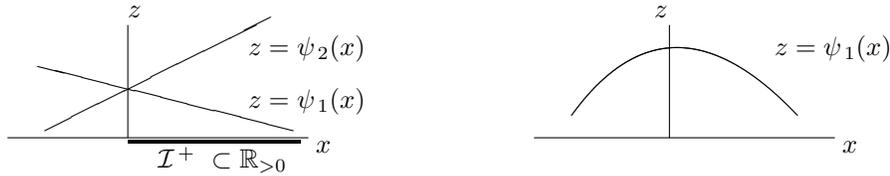

We focus on the low temperature phase (symmetry broken phase), 
since the system in the
high temperature phase is equivalent to systems with no-phase transitions and
has been addressed\,\cite{Goto2015JMP}.  
To discuss system in Fig.\,\ref{Husimi-partial-hysteresis1-picture},  
label branches of the $2$-valued function of $x$ 
as in Fig.\,\ref{Husimi-legndre-picture-labels2}
(left, low temperature phase),
where the function $\psi_{\,J_{0}}$ does not depend on $y$
on the Legendre submanifold 
due to \fr{Husimi-psi-y-derivative} with \fr{Husimi-psi-x-derivative}.  
Then, on the Legendre submanifold generated by 
$\psi_{\,J_{0}}$,
the abbreviation $\psi_{\,\mu}(x)=\psi_{\,J_{0}}(x,y_{\,\mu}^{\,*})$ is introduced
for each $\mu$. 
  The region $\cI^{\,+}\subset\mbbR_{\,>0}$ is defined such that there
  are two (single-valued) functions, in particular
  $\psi_{\,1}$ and $\psi_{\,2}$ are labeled such that  
  $\psi_{\,1}(x)<\psi_{\,2}(x)$, $(x\in\cI^{\,+})$. That is, 
  $\cI^{\,+}:=\{\, x\in\mbbR_{\,>0}\,|\,\psi_{\,1}(x)<\psi_{\,2}(x)\,\}$,
  (see Figs. \ref{picture-outline-pruned-Legendre}
    and \ref{Husimi-legndre-picture-labels2}).   
In the high temperature phase, the (single-valued) function appears. 
Then, decompose the subset of the 
Legendre submanifold 
$\phi\cA_{\,\psi_{\,J_{0}}}$ into the ones with $\cI^{\,+}$
$$
\phi\cA_{\,\mu}^{\,\cI^{\,+}}
=\left\{\ (x,y,z)\in\cC 
\ \bigg|\ y=-\,\frac{\dr\psi_{\,\mu}}{\dr x} ,\ z=\psi_{\,\mu}(x),\quad
x\in\cI^{\,+} 
\ \right\},\quad \mu=1,2.
$$

One then can show the main theorem of this paper as below.
Notice that no explicit expression of $\psi_{\,\mu}$ defined on $\cI^{\,+}$ 
is needed for each $\mu$. 
Other different Theorems closely related to this main theorem
  are stated in Appendix\,\ref{section-appendix}. 

\begin{Theorem}
  \label{claim-cubic-contact-Hamiltonian3}
  (satable segments of the hysteresis curve in the symmetry broken phase).
  On the thermodynamic phase space
  for the Husimi-Temperley model, 
  choose a contact Hamiltonian $h$ as 
\beq
  h(x,z)
  = -\psi_{\,0}(x)(z-\psi_{\,1}(x))(z-\psi_{\,2}(x))^{\,2},
\label{wave-fronts-as-attractor-contact-hamiltonian3}
  \eeq
  where
  $\psi_{\,0}$  is an arbitrary function of $x$ such that  $\psi_{\,0}(x)>0$.
Then the following hold. 
  \begin{enumerate}
\item    
  The space $\phi\cA_{\,1}^{\,\cI^{\,+}}$ is asymptotically stable in
  $\cD_{\,1}^{\,+}$, 
  where  
  $$
  \cD_{\,1}^{\,+}
  =  \{\ (x,y,z) 
  \ | \ x\in\cI^{\,+},\ z<\psi_{\,2}(x)\  
  \}.
  $$
\item
  The space $\phi\cA_{\,2}^{\,\cI^{\,+}}$ is asymptotically stable in
  $\cD_{\,2}^{\,+}$, 
  where  
  $$
  \cD_{\,2}^{\,+}
  =  \{\ (x,y,z) 
  \ | \ x\in\cI^{\,+},\ \psi_{\,2}(x)\leq z \  
  \}.
  $$
  
  \end{enumerate}
\end{Theorem}
\begin{Proof}
  Our strategy for proving this is to show the existence of 
  Lyapunov functions\,\cite{smale} for  
  the dynamical system, where this dynamical system is obtained from
  substituting the contact Hamiltonian 
\fr{wave-fronts-as-attractor-contact-hamiltonian3} into 
\fr{contact-hamiltonian-vector-field-component}. 
The details are as follows.

First, a point of departure for this proof is to express 
the explicit form of the
dynamical system written in terms of the coordinates $(x,y,z)$.  
From \fr{contact-hamiltonian-vector-field-component}
and \fr{wave-fronts-as-attractor-contact-hamiltonian3}, the
dynamical system is explicitly written as   
\beqa
\dot{x}
&=&0,
\label{wave-fronts-as-attractor-x3}\\
\dot{y}
&=&\frac{\dr \psi_{\,0}}{\dr x}(z-\psi_{\,1})(z-\psi_{\,2})^{\,2}
-\psi_{\,0}\left(y+\frac{\dr\psi_{\,1}}{\dr x}\right)(z-\psi_{\,2})^{\,2}
-2\psi_{\,0}\left(y+\frac{\dr\psi_{\,2}}{\dr x}\right)(z-\psi_{\,1})(z-\psi_{\,2}),
\label{wave-fronts-as-attractor-y3}\\
\dot{z}
&=&h
=-\, \psi_{\,0}(x)(z-\psi_{\,1}(x))(z-\psi_{\,2}(x))^{\,2}.
\label{wave-fronts-as-attractor-z3}
\eeqa
The next step is to find fixed point sets. 
From 
$$
\dot{x}|_{\,\phi\cA_{\mu}^{\cI^{+}}}
=0,\quad
\dot{y}|_{\,\phi\cA_{\mu}^{\cI^{+}}}
=0,\quad
\dot{z}|_{\,\phi\cA_{\mu}^{\cI^{+}}}
=0,\qquad \mu=1,2
$$
one has that $\phi\cA_{\,\mu}^{\cI^{+}}\subset\cC$, $(\mu=1,2)$
forms a fixed point set for each $\mu$. 
Here a phase portrait of the dynamical system is roughly discussed.
It follows from 
\fr{wave-fronts-as-attractor-x3} that $x$ is constant in time, and
thus $\psi_{\,\mu}(x)$ does not depend on time.

Third, to prove the theorem, 
Lyapunov functions are constructed\,\cite{smale}.
Define the functions $V_{\,1}$ on $\cD_{\,1}^{\,+}$,
and $V_{\,2}$ on $\cD_{\,2}^{\,+}$, 
such that 
\beqa
V_{\,1}(x,z)
&=&\frac{1}{2}(z-\psi_{\,1}(x))^{\,2},\qquad (x,y,z)\in\cD_{\,1}^{\,+},
\non\\
V_{\,2}(x,z)
&=&z-\psi_{\,2}(x),\qquad (x,y,z)\in\cD_{\,2}^{\,+}.
\non
\eeqa
Then, differentiation of $V_{1}$ and that of $V_{2}$ yield the following. 
\begin{itemize}
\item
  It follows that
  $$  
V_{\,1}(x,z)
\geq 0,\quad
\frac{\dr V_{\,1}}{\dr t}
=(z-\psi_{\,1})h(x,z)
=-\,\psi_{\,0}(x)(z-\psi_{\,1})^{\,2}(z-\psi_{\,2})^{\,2}
\leq 0,\qquad (x,y,z)\in\cD_{\,1}^{\,+},
$$
where the equality holds on the fixed point set 
$\phi\cA_{\,1}^{\,\cI^{\,+}}$.
Hence $V_{\,1}$ is a Lyapunov function on $\cD_{\,1}^{\,+}$.
\item
  It follows that
  $$
V_{\,2}(x,z)
\geq 0,\quad
\frac{\dr V_{\,2}}{\dr t}
=h(x,z)
=-\,\psi_{\,0}(x)(z-\psi_{\,1})(z-\psi_{\,2})^{\,2}
\leq 0,\qquad (x,y,z)\in\cD_{\,2}^{\,+},
$$
where the equality holds on the fixed point set 
$\phi\cA_{\,2}^{\,\cI^{\,+}}$.
Hence $V_{\,2}$ is a Lyapunov function on $\cD_{\,2}^{\,+}$.
\end{itemize}
According to the theorem of Lyapunov, one completes the proof.
\qed
\end{Proof}

Theorem\,\ref{claim-cubic-contact-Hamiltonian3} shows that  
the proposed contact Hamiltonian vector field is such that
the pruned segments of the projected Legendre submanifold are stable 
in regions of a contact manifold. 
The global behavior for $z$ is understood from
Fig.\,\ref{Husimi-legndre-picture-flow-z3}.
One then deduces from Fig.\,\ref{Husimi-legndre-picture-flow-z3}
that, given $x$, 
$\lim_{t\to\infty}z(t)=\psi_{\,1}(x)$ in $\cD_{\,1}^{\,+}$, and that 
$\lim_{t\to\infty}z(t)=\psi_{\,2}(x)$ in $\cD_{\,2}^{\,+}$. 
If initial states are near $\phi\cA_{\,2}^{\,+}$ in $\cD_{\,1}^{\,+}$,
then the integral curves connect metastable states and the most stable states.
Meanwhile, in the case where there is only one (single-valued) function $\psi$ of $x$ 
defined on a region in $\mbbR$,  
one can find a contact Hamiltonian system 
such that the projection of the Legendre submanifold is 
stable as has been argued in Refs.\,\cite{Goto2015JMP,Entov2021}.

\begin{figure}
\begin{picture}(120,29)
\unitlength 1mm
\put(47,26){$\dot{z}$}
\put(111,10){$z$}
\put(52,21){$\dot{z}=h(x,z)$}
\put(51,11){\line(0,1){15}}
\put(51,11){\line(1,0){58}}
\put(47,10){$0$}
\qbezier(50,20)(58,5)(75,10)
\qbezier(75,10)(90,14)(105,2)
\put(54,3){$z=\psi_{\,1}(x)$}
\put(77,3){$z=\psi_{\,2}(x)$}
\put(58.5,5){\vector(0,1){5}}
\put(82,5){\vector(0,1){5}}
%
\linethickness{0.5mm}
\end{picture}
\caption{ 
Phase space of
  the dynamical system consisting of $\dot{z}=h(x,z)$ and $\dot{x}=0$
  (\fr{wave-fronts-as-attractor-z3} and 
  \fr{wave-fronts-as-attractor-x3}, respectively).
 From $\dot{z}=h$, it follows that the zeros of
  $h$ are the fixed points.
 From $h$ in \fr{wave-fronts-as-attractor-contact-hamiltonian3}
  its zeros are the set $z=\psi_{\,1}(x)$ and the set $z=\psi_{\,2}(x)$. 
  In addition, from the sign of $h$, it follows that the set  $z=\psi_{\,1}(x)$  
  is stable in $\cD_{\,1}^{\,+}$ and that the set $z=\psi_{\,2}(x)$ is stable in $\cD_{\,2}^{\,+}$.
  }
\label{Husimi-legndre-picture-flow-z3}
\end{figure}

Physically, 
the contact Hamiltonian vector field with
\fr{wave-fronts-as-attractor-contact-hamiltonian3} 
expresses the dynamical
process towards equilibrium states when initial states are not at equilibrium.  

To grasp local flow around the fixed point set 
$(y_{\,1},z_{\,1})=(-\psi_{\,1}^{\,\prime},\psi_{\,1})$, the 
integral curves of the linearized equations are shown below.
For the point  $(y_{\,1},z_{\,1})=(- \psi_{\,1}^{\,\prime},\psi_{\,1})$,
introduce $Y_{\,1}$ and $Z_{\,1}$ such that 
$$
y(t)
=- \psi_{\,1}^{\,\prime}(x)+Y_{\,1}(t),\quad
z(t)
=\psi_{\,1}(x)+Z_{\,1}(t),\quad\mbox{where}\quad
\psi_{\,1}^{\,\prime}(x)
:=\frac{\dr \psi_{\,1}}{\dr x}(x),
$$
which yield linearized equations. For ease of notation, introduce 
\beq
\psi_{\,21}(x)
=\psi_{\,2}(x)-\psi_{\,1}(x)\ 
>0,
\label{linearized-Z-2-branches-h3}
\eeq
for each point $x$. 
Then the linearized equations are obtained as 
$$
\dot{Z}_{\,1}
=-\,\underbrace{\psi_{\,0}\,\psi_{\,21}^{\ 2}}_{>\ 0}\, Z_{\,1},
\qquad
\dot{Y}_{\,1}
=-\underbrace{\psi_{\,0}\,\psi_{\,21}^{\ 2}}_{>\ 0}\, Y_{\,1}
+(\,\psi_{\,0}^{\,\prime}\,\psi_{\,21}^{\ 2}
+\psi_{\,0}\,\psi_{\,21}\,\psi_{\,21}^{\,\prime}\,)\,Z_{\,1}, 
$$
where $\psi_{\,21}^{\,\prime}(x)
=\dr\psi_{\,21}/\dr x$ and 
$\psi_{\,0}^{\,\prime}=\dr\psi_{\,0}/\dr x$ are constants in time.  
To solve this linear system of equations, letting $c_{\,1}$ 
and $d_{\,1}$ be the constants such that
$$
c_{\,1}
=\psi_{\,0}\,\psi_{\,21}^{\ 2},\qquad
d_{\,1}
=\psi_{\,0}^{\,\prime}\,\psi_{\,21}^{\ 2}
+\psi_{\,0}\,\psi_{\,21}\,\psi_{\,21}^{\,\prime},
$$
one can write 
$$
\dot{Z}_{\,1}
=-\,c_{\,1} Z_{\,1},\quad
\dot{Y}_{\,1}
=-\,c_{\,1} Y_{\,1}+d_{\,1}Z_{\,1},
$$
The solution of this system is 
$$
Z_{\,1}(t)
=Z_{\,1}(0)\,\e^{\,-c_{1}\,t},\quad
Y_{\,1}(t)
=(\,Y_{\,1}(0)+d_{\,1}Z_{\,1}(0)\, t\,)\,\e^{\,- c_{1}\,t}.
$$
From this solution and the inequality $c_{\,1}>0$, 
one has that the fixed point set  
$\phi\cA_{\,1}^{\cI^{\,+}}$  
is linearly stable. 
Observe from
\fr{linearized-Z-2-branches-h3} 
that the strength of stability is large when the value  
$c_{\,1}=\psi_{\,0}\,\psi_{\,21}^{\,2}$ is large. 
The condition when $\psi_{\,21}(x)$ is large can be read off  
from Fig.\,\ref{Husimi-partial-hysteresis1-picture}.
The value $\psi_{\,21}(x)$ is small near the critical point, and  
it is large far from the critical point. 

So far the phase space  $\cD_{\,1}^{\,+}\cup\cD_{\,2}^{\,+}$
of the dynamical system is focused,  
and then a similar claim can be stated for the
region $\cD_{\,1}^{\,-}\cup \cD_{\,2}^{\,-}$, where 
$\cD_{\,1}^{\,-}$ and $\cD_{\,2}^{\,-}$ are defined with some
$\cI^{\,-}\subset\mbbR_{<0}$.
To state such a claim, introduce some notations as follows. 
  Similar to the sets 
    $\cD_{\,1}^{\,+}=\{(x,y,z)\,|\,x\in\cI^{\,+},z<\psi_{\,2}(x)\}$, and 
  $\cI^{\,+}=\{\, x\in\mbbR_{\,>0}\,|\,\psi_{\,1}(x)<\psi_{\,2}(x)\,\} \subset\mbbR$,
  introduce
  $$
  \cD_{\,1}^{\,-}
  =\{(x,y,z)\ \,|\,x\in\cI^{\,-},z<\psi_{\,2}(x)\},
  $$
  where $\cI^{\,-}$  
  has been defined 
  in the caption to Fig.\,\ref{picture-outline-pruned-Legendre} as
  $$
  \cI^{\,-}
  =\{\, x\in\mbbR_{\,<0}\,|\,\psi_{\,1}(x)<\psi_{\,2}(x)\,\} \subset\mbbR.
  $$

By combining these and refining it, one has the following.
\begin{Corollary}
  \label{fact-combining-the-contact-Hamiltonian-system-h3}
  (reconstruction of the stability of the hysteresis and
  pseudo-free energy curves as the Legendre submanifold).  
  Consider the system shown in   Fig.\,\ref{Husimi-Legendre-hysteresis-picture}.
  In the joined region 
  $\cD_{\,1}^{\,+}\cup \cD_{\,1}^{\,-}$ in the contact manifold $\cC$,  
  one has the contact Hamiltonian vector fields where the undirected curves
  $\ol{\mathrm{0-i}}$ and  $\ol{\mathrm{v-vi}}$ 
  are stable fixed point sets for $z<\psi_{\,2}$, 
  and
  $\ol{\mathrm{i-ii}}$ and  $\ol{\mathrm{iv-v}}$ are
  stable fixed point sets of
  the curve for $\psi_{\,2}\leq z$.  
\end{Corollary}  
Notice that the  $(y,z)$-plane at $x=0$  
has been removed from $\cC$ in Corollary\,\ref{fact-combining-the-contact-Hamiltonian-system-h3}. 
On this removed plane, the
  double-valued function becomes a single valued function, and thus 
  the present contact Hamiltonian is not relevant. 
In addition, even if the segment $\ol{\mathrm{ii-iv}}$ is unremoved in 
Fig.\,\ref{Husimi-Legendre-hysteresis-picture}, flows of the contact Hamiltonian 
system are not affected by the existence of the segment $\ol{\mathrm{ii-iv}}$. 
  
\begin{figure}[htb]
\centering
\includegraphics[width=6.7cm]{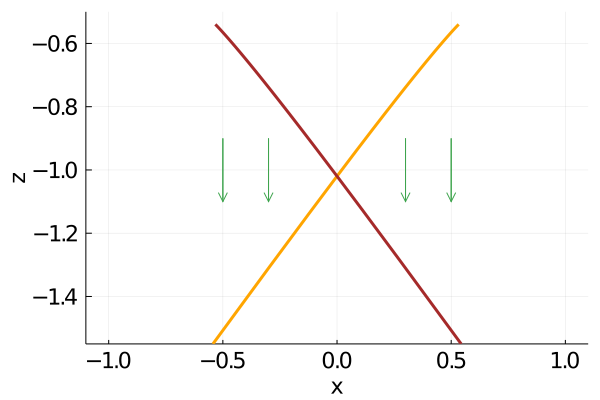}
\includegraphics[width=6.7cm]{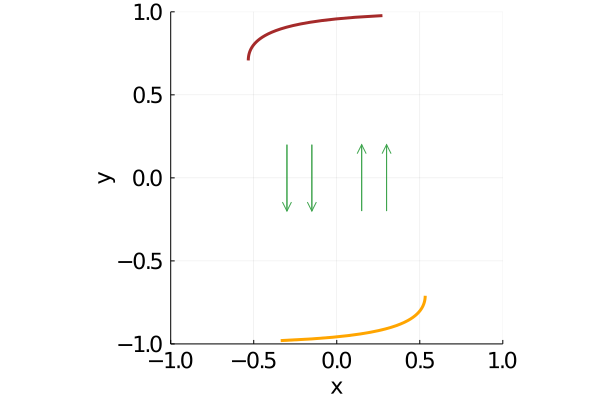}
\caption{
  Projected contact Hamiltonian vector fields 
  stated in
  Corollary\,\ref{fact-combining-the-contact-Hamiltonian-system-h3},
  where the values of $\beta$ and $J_{\,0}$ were chosen to express
  a thermodynamic phase space being 
  far from the critical point in the low temperature phase 
  ($J_{\,0}=1.0$ 
  in the dimensionless variable constructed from 
  $\beta=1.0$ and $J_{\,0}=1.0$ in the dimensional variables). 
The pruned segments of the projections of the Legendre submanifold are  
stable and unstable fixed point sets. 
(Left) The $(x,z)$-plane. (Right) The $(x,y)$-plane. }
\label{Husimi-partial-hysteresis1-vector-field-picture}
\end{figure}

In Fig.\,\ref{Husimi-partial-hysteresis1-vector-field-picture},
the projected contact Hamiltonian vector fields 
stated in 
Corollary\,\ref{fact-combining-the-contact-Hamiltonian-system-h3} are
shown. From this corollary, one has the following.
\begin{Remark}\label{remark-asymptotic-phase-portrait} 
  \begin{enumerate}
\item 
  The cusp of the shape $\wedge$ on the $(x,z)$-plane is obtained 
  in the long-time limit of the time-development 
    of the contact Hamiltonian system,  
  where such a shape  
  is expected to be observed in experiments under perturbation. 
  \item
The kink structure of the shape  
$\step$ on the $(x,y)$-plane is obtained 
in the long-time limit of the time-development of the contact 
  Hamiltonian system,  
where such a kink shape 
is  
expected to be observed in experiments under perturbation. 
\end{enumerate}
\end{Remark}

\section{Discussions and conclusions}
\label{Conclusions}

This paper offers a contact geometric
approach to thermodynamic systems that exhibit 
a phase transition. One key in this paper has been that 
the set of metastable, unstable, and the most stable 
equilibrium states
is identified with 
a Legendre submanifold whose projections form  multi-valued functions. 
As the main theorem of this paper unstable and stable 
segments of a hysteresis curve have been 
described as unstable and stable fixed
point sets for a contact Hamiltonian vector field.
Simultaneously the pseudo-free energy curve has also been 
  described similarly, where this simultaneity is ascribed to the different
  projections of the unique Legendre submanifold.
  On this $1$-dimensional Legendre submanifold there is no singularity
  even in the symmetry broken phase. Meanwhile there are singularities on the
  $2$-dimensional plane as the result of the projection. 
Although
these calculations have been for the so-called
Husimi-Temperley model, 
calculations for other models and those for the 
present model are expected to be similar. 
The series of calculations has been summarized as a
procedure, and then this  
procedure has
been summarized in Introduction of this paper. 
A significance of this study is to
provide a unified geometric manner
in which a contact Hamiltonian and
a single Legendre submanifold together with an associated    
pruning process 
lead to various notions and tools in thermodynamics.  
Such notions and tools are  non-most-stable and most stable equilibrium
states,
hysteresis and pseudo-free energy curves, and 
rules similar to the Maxwell construction and
the convexification.

There remain unsolved problems that have not been addressed in this paper.
They include the following: 
\begin{itemize}
\item
derivation of 
the contact Hamiltonian vector field from a dynamical system that describes 
microscopic spins\,\cite{Mori2010},

\item
application of 
  the present approach to various statistical mechanical models,  
thermodynamic systems, and electric circuits\,\cite{Goto2016},  

\item 
extension of 
the present or similar analysis to high-dimensional systems,
rather than the present $3$-dimensional contact manifold,
\item
application of 
Legendre 
singularity and cobordism theories intensively
to thermodynamic systems\,\cite{Aicardi2001,Arnold-Givental},  
\item
clarification of 
the relation between a contact version of
the {\it (graph) selector}\,\cite{Gabriel2003,Limouzineau2016} 
and the pruning introduced in this paper.
  \end{itemize}
By addressing these, it is expected that a relevant and sophisticated
geometric methodology will be established for dealing with 
various intricate systems and    
critical phenomena.

 \subsection*{Data Availability Statement}
 In this paper various numerical figures have been drawn 
 with the Julia  language on a personal computer. 
To draw these figures,  
numerical plots have been made with 
 elementary functions, including $\log$, $\tanh$, and so on.
 Hence there is no raw data, and these figures are easily reproduced. 
 For this reason, data sharing is not applicable to this article
 as no new data were created or analyzed in this study.

 \section*{Acknowledgment}
 The author was partially supported by
 JSPS (KAKENHI) grant number JP19K03635, and thanks   
 Lenonid Polterovich for discussions on applications of contact geometry to
 nonequilibrium statistical mechanics. 
   In addition the author thanks Minoru Koga for giving various
   suggestions and fruitful discussions on this study.  

\appendix
\renewcommand{\theequation}{A.\arabic{equation}}
\renewcommand{\fr}[1]{(\ref{#1})}
\setcounter{equation}{0}
 \section{Appendix:\ Various contact Hamiltonian systems}
\label{section-appendix}
For the sake of completeness, 
a system with the unpruned projection of the Legendre submanifold is argued  
in this section. 
  In addition, a system similar to that in
  Theorem\,\ref{claim-cubic-contact-Hamiltonian3} is considered, and then  
  it is shown that in $z>\psi_{\,2}$ the Legendre submanifold
  $\phi\cA_{\,2}^{\,\cI^{\,+}}$ forms a fixed point set but 
  is unstable. 
Several contact Hamiltonian systems are considered in the main text and
Appendix in this paper. 
They are summarized in Table\,\ref{table-various-contact-Hamiltonian}.
From a physical viewpoint, in Theorem\,\ref{claim-cubic-contact-Hamiltonian},
one unsatisfactory feature
is that $\phi \cA_{\,3}^{\,\cI^{+}}$ is stable, and another one is that
the (meta-)stability of $\phi\cA_{\,2}^{\,+}$ is lost. 
In addition, in
Theorem\,\ref{claim-cubic-contact-Hamiltonian2}, one unsatisfactory feature 
is that the (meta-)stability of $\phi\cA_{\,2}^{\,\cI^{+}}$ is lost. 
\begin{table}[htb]
  \begin{center}
    \begin{tabular}{|c|l|c|l|}
      \hline
      Theorem & Contact Hamiltonian $h$ & Graph of $h$ & Stable fixed point sets\\
      \hline
      \ref{claim-cubic-contact-Hamiltonian3}
      & $h=-\psi_{\,0}(z-\psi_{\,1})(z-\psi_{\,2})^{\,2}$
      & Fig.\,\ref{Husimi-legndre-picture-flow-z3}  
      & $\phi\cA_{\,1}^{\,\cI^{+}}$ in $\cD_{\,1}^{\,+}$, and 
      $\phi\cA_{\,2}^{\,\cI^{+}}$ in $\cD_{\,2}^{\,+}$
      \\
      \ref{claim-cubic-contact-Hamiltonian}
      &  $h=-\psi_{\,0}(z-\psi_{\,1})(z-\psi_{\,2})(z-\psi_{\,3})$
      & Fig.\,\ref{Husimi-legndre-picture-flow-z} 
      & $\phi\cA_{\,1}^{\,\cI^{+}}$ in $\cD_{\,1}^{\,+}$, and 
      $\phi\cA_{\,3}^{\,\cI^{+}}$ in $\cD_{\,3}^{\,+}$
      \\
      \ref{claim-cubic-contact-Hamiltonian2}
      & $h=\ \ \psi_{\,0}(z-\psi_{\,1})(z-\psi_{\,2})$
      & Fig.\,\ref{Husimi-legndre-picture-flow-z2} 
      & $\phi\cA_{\,1}^{\,\cI^{+}}$  in $\cD_{\,1}^{\,+}$
      \\
      \hline
    \end{tabular}
    \end{center}
  \caption{Various contact Hamiltonian systems}
  \label{table-various-contact-Hamiltonian}
\end{table}

\subsection{Three branch system}
To discuss system in Fig.\,\ref{Husimi-wave-front-picture}, 
label branches of the $3$-valued function of $x$  
as in Fig.\,\ref{Husimi-legndre-picture-labels} (left, low temperature phase),
where the function $\psi_{\,J_{0}}$ does not depend on $y$
on the Legendre submanifold 
due to \fr{Husimi-psi-y-derivative} with \fr{Husimi-psi-x-derivative}. 
Then, as in the case of 
Section\,\ref{section-nonequilibrium}, 
on the Legendre submanifold generated by $\psi_{\,J_{0}}$, 
the abbreviation $\psi_{\,\mu}(x)=\psi_{\,J_{0}}(x,y_{\,\mu}^{\,*})$ is
introduced for each $\mu$. 
In the region $\cI^{\,+}\subset\mbbR$, there are three (single-valued) 
functions $\psi_{\,1},\psi_{\,2},\psi_{\,3}$ labeled such that  
$\psi_{\,1}(x)<\psi_{\,2}(x)<\psi_{\,3}(x)$, $(x\in\cI^{\,+})$. 
In the high temperature phase, the (single-valued) function appears.

\begin{figure}
\begin{picture}(120,60)
\unitlength 1mm
\put(36,20){$z$}
\put(61,2){$x$}
\put(58,22){$z=\psi_{\,3}(x)$}
\put(52,15){$z=\psi_{\,2}(x)$}
\put(52,8){$z=\psi_{\,1}(x)$}
\put(36,4){\line(0,1){15}}
\put(20,4){\line(1,0){40}}
\put(40,0){$\cI^{\,+}\, \subset\mbbR$}
\qbezier(25,16)(38,13)(55,22)
\put(58,5){\line(-4,1){34}}
\put(25,5){\line(2,1){30}}
\put(108,20){$z$}
\put(131,2){$x$}
\put(122,15){$z=\psi_{\,1}(x)$}
\put(108,4){\line(0,1){15}}
\put(90,4){\line(1,0){40}}
\qbezier(95,7)(108,25)(125,7)
\linethickness{0.5mm}
\put(36,3.5){\line(1,0){23}}
\end{picture}
\caption{Wave front. (Left) Low temperature phase.
The label $\mu$ for $\psi_{\,\mu}$ is  chosen so that
  $\psi_{\,1}(x)<\psi_{\,2}(x)<\psi_{\,3}(x)$.  
  (Right) High temperature phase, the (single-valued) function appears.}
\label{Husimi-legndre-picture-labels}
\end{figure}

We focus on the low temperature phase again as in the case of the $2$-valued
function. 
To discuss the low temperature phase, decompose the subset of the
Legendre submanifold 
$\phi\cA_{\,\psi_{\,J_{0}}}$ into the ones with $\cI^{\,+}$
$$
\phi\cA_{\,\mu}^{\,\cI^{+}}
=\left\{\ (x,y,z) 
\ \bigg|\ y=-\,\frac{\dr\psi_{\,\mu}}{\dr x} ,\ z=\psi_{\,\mu}(x),\quad
x\in\cI^{\,+} 
\ \right\},\quad \mu=1,2,3.
$$
 
One can show the following theorem. Notice,
as well as in Theorem\,\ref{claim-cubic-contact-Hamiltonian3}, 
that no explicit expression of $\psi_{\,\mu}$ defined on $\cI^{\,+}$
is needed for each $\mu$. 
\begin{Theorem}
  \label{claim-cubic-contact-Hamiltonian}
(stable and unstable segments of the hysteresis curve in the symmetry broken phase).  
  On the thermodynamic phase space
  for the Husimi-Temperley model,  
  choose a contact Hamiltonian $h$ as 
\beq
  h(x,z)
  =- \psi_{\,0}(x)(z-\psi_{\,1}(x))(z-\psi_{\,2}(x))(z-\psi_{\,3}(x)).
\label{wave-fronts-as-attractor-contact-hamiltonian}
  \eeq
  where
  $\psi_{\,0}$  is an arbitrary function of $x$ such that  $\psi_{\,0}(x)>0$.
Then the following hold. 
  \begin{enumerate}
\item    
  The space $\phi\cA_{\,1}^{\,\cI^{+}}$
  is asymptotically stable in $\cD_{\,1}^{+}$, 
  where  
  $$
  \cD_{\,1}^{\,+}
  =  \{\ (x,y,z) 
  \ | \
x\in\cI^{\,+},\ z<\psi_{\,2}(x)\  
  \}.
  $$
\item
  The space $\phi\cA_{\,3}^{\,\cI^{+}}$
  is asymptotically stable in $\cD_{\,3}^{+}$, 
  where 
  $$
  \cD_{\,3}^{\,+}
  =  \{\ (x,y,z) 
  \ | \
x\in \cI^{\,+},\ z>\psi_{\,2}(x)\  
  \}.
  $$
  \end{enumerate}
\end{Theorem}
\begin{Proof}
  Our strategy for proving this is to show the existence of a
  Lyapunov function\,\cite{smale} for  
  the dynamical system obtained from
  substituting the contact Hamiltonian 
\fr{wave-fronts-as-attractor-contact-hamiltonian} into 
\fr{contact-hamiltonian-vector-field-component}. 
The details are as follows.

First, a point of departure for this proof is to express
the explicit form of the
dynamical system written in terms of the coordinates $(x,y,z)$.  
From \fr{contact-hamiltonian-vector-field-component}
and \fr{wave-fronts-as-attractor-contact-hamiltonian}, the
dynamical system is explicitly written as  
\beqa
\dot{x}
&=&0,
\label{wave-fronts-as-attractor-x}\\
\dot{y}
&=&\frac{\dr \psi_{\,0}}{\dr x}(z-\psi_{\,1})(z-\psi_{\,2})(z-\psi_{\,3})
-\psi_{\,0}\left(y+\frac{\dr\psi_{\,1}}{\dr x}\right)(z-\psi_{\,2})(z-\psi_{\,3})
\non\\
&&-\psi_{\,0}\left(y+\frac{\dr\psi_{\,2}}{\dr x}\right)(z-\psi_{\,1})(z-\psi_{\,3})-\psi_{\,0}\left(y+\frac{\dr\psi_{\,3}}{\dr x}\right)(z-\psi_{\,1})(z-\psi_{\,2})
\label{wave-fronts-as-attractor-y}\\
\dot{z}
&=&h
=-\, \psi_{\,0}(x)(z-\psi_{\,1}(x))(z-\psi_{\,2}(x))(z-\psi_{\,3}(x)).
\label{wave-fronts-as-attractor-z}
\eeqa
The next step is to find fixed point sets. 
From 
$$
\dot{x}|_{\,\phi\cA_{\mu}^{\,\cI^{+}}}
=0,\quad
\dot{y}|_{\,\phi\cA_{\mu}^{\,\cI{+}}}
=0,\quad
\dot{z}|_{\,\phi\cA_{\mu}^{\,\cI^{+}}}
=0.\qquad \mu=1,2,3
$$
one has that $\phi\cA_{\,\mu}^{\,\cI^{+}}\subset\cC$, $(\mu=1,2,3)$
forms a fixed point set for each $\mu$. 
Here a phase portrait of the dynamical system is roughly discussed.
It follows from 
\fr{wave-fronts-as-attractor-x} that $x$ is constant in time, and
thus $\psi_{\,\mu}(x)$ does not depend on time.

Third, to prove the theorem, 
Lyapunov functions are constructed \,\cite{smale}. 
Define the functions $V_{\,1}$ in $\cD_{\,1}^{\,+}$ 
and $V_{\,3}$ in $\cD_{\,3}^{\,+}$ 
such that 
\beqa
V_{\,1}(x,z)
&=&\frac{1}{2}(z-\psi_{\,1}(x))^{\,2},\qquad (x,y,z)
\in\cD_{\,1}^{\,+}
\non\\
V_{\,3}(x,z)
&=&\frac{1}{2}(z-\psi_{\,3}(x))^{\,2},\qquad (x,y,z)
\in\cD_{\,3}^{\,+}.
\non
\eeqa
Then,
differentiation of $V_{\,1}$ and that of $V_{\,3}$ yield the following. 
\begin{itemize}
  \item
In $\cD_{\,1}^{\,+}$,  it follows that 
    $$
V_{\,1}(x,z)
\geq 0,\quad
\frac{\dr V_{\,1}}{\dr t}
=(z-\psi_{\,1})h(x,z)
=-\psi_{\,0}(z-\psi_{\,1})^{\,2}(z-\psi_{\,2})(z-\psi_{\,3})
\leq 0,\qquad (x,y,z)\in\cD_{\,1}^{\,+}, 
$$
where the equality holds on the fixed point set 
$\phi\cA_{\,1}^{\,\cI^{+}}$.
Hence $V_{\,1}$ is a Lyapunov function in
$\cD_{\,1}^{\,+}$. 
\item
In $\cD_{\,3}^{\,+}$, it follows that   
  $$
V_{\,3}(x,z)
\geq 0,\quad
\frac{\dr V_{\,3}}{\dr t}
=(z-\psi_{\,3})h(x,z)
=-\psi_{\,0}(z-\psi_{\,1})(z-\psi_{\,2})(z-\psi_{\,3})^{\,2}
\leq 0,\qquad (x,y,z)
\in\cD_{\,3}^{\,+}, 
$$
where equality holds on the fixed point set 
$\phi\cA_{\,3}^{\,\cI^{+}}$.
Hence $V_{\,3}$ is a Lyapunov function in $\cD_{\,3}^{\,+}$. 
\end{itemize}
According to the theorem of Lyapunov, one completes the proof.
\qed
\end{Proof}

The global behavior for $z$ is understood from
Fig.\,\ref{Husimi-legndre-picture-flow-z}.
One then deduces from Fig.\,\ref{Husimi-legndre-picture-flow-z}
that, given $x$, 
$\lim_{t\to\infty}z(t)=\psi_{\,1}(x)$ in $\cD_{\,1}^{\,+}$, and
that  
$\lim_{t\to\infty}z(t)=\psi_{\,3}(x)$ in $\cD_{\,3}^{\,+}$. 

\begin{figure}
\begin{picture}(120,29)
\unitlength 1mm
\put(47,26){$\dot{z}$}
\put(111,10){$z$}
\put(92,18){$\dot{z}=h(x,z)$}
\put(51,11){\line(0,1){15}}
\put(51,11){\line(1,0){58}}
\put(47,10){$0$}
\qbezier(50,20)(60,0)(75,10)
\qbezier(75,10)(90,20)(105,10)
\put(51,3){$z=\psi_{\,1}$}
\put(69,17){$z=\psi_{\,2}(x)$}
\put(99,3){$z=\psi_{\,3}(x)$}
\put(56,5){\vector(0,1){5}}
\put(77,15.7){\vector(0,-1){5}}
\put(103,5){\vector(0,1){5}}
\linethickness{0.5mm}
\end{picture}
\caption{Phase space of
  the dynamical system consisting of $\dot{z}=h(x,z)$ and $\dot{x}=0$
  (\fr{wave-fronts-as-attractor-z} and 
  \fr{wave-fronts-as-attractor-x}).
  From $\dot{z}=h$, it follows that the zeros of
  $h$ are the fixed points.
  From $h$ in \fr{wave-fronts-as-attractor-contact-hamiltonian}
  its zeros are the set $z=\psi_{\,1}(x)$, $z=\psi_{\,2}(x)$ and 
  the set $z=\psi_{\,3}(x)$.
  In addition, from the sign of $h$, the set $z=\psi_{\,1}(x)$
  and the set $z=\psi_{\,3}(x)$ are stable in some domains. 
  }
\label{Husimi-legndre-picture-flow-z}
\end{figure}

To elucidate the behavior of the contact Hamiltonian vector field in
Theorem\,\ref{claim-cubic-contact-Hamiltonian} on the
lower dimensional spaces that have been used for the projections, see
Fig.\,\ref{Husimi-Legendre-hysteresis-picture}.
This Theorem states that $\phi\cA_{\,1}^{\,\cI^{+}}$ and 
$\phi\cA_{\,3}^{\,\cI^{+}}$ are stable.
This is equivalent to say that the part of Legendre curves 
$\ol{\mathrm{v-vi}}$ and $\ol{\mathrm{ii-iii}}$ are stable in some domains. 
Although it is not immediately clear how the stability of the 
curve $\ol{\mathrm{ii-iii}}$ plays a role in physical context, the role of
stability of the curve $\ol{\mathrm{v-vi}}$ is clear. That stability
for $\ol{\mathrm{v-vi}}$ is consistent with the thermodynamic stability.

To grasp local flow around the fixed point sets, 
integral curves of the linearized equations are shown below.
For the point  $(y_{\mu},z_{\mu})=(- \psi_{\,\mu}^{\,\prime},\psi_{\,\mu})$,
introduce $Y_{\,\mu}$ and $Z_{\,\mu}$
such that 
$$
y(t)
=- \psi_{\,\mu}^{\,\prime}(x)+Y_{\,\mu}(t),\quad
z(t) 
=\psi_{\,\mu}(x)+Z_{\,\mu}(t),\quad\mbox{where}\quad
\psi_{\,\mu}^{\,\prime}(x)
:=\frac{\dr \psi_{\,\mu}}{\dr x}(x),
\quad\mu=1,2,3
$$
which yield linearized equations. For ease of notation, introduce 
\beq
\psi_{\,\mu\mu^{\prime}}
=\psi_{\,\mu}(x)-\psi_{\,\mu^{\prime}}(x),
\quad\mu,\mu^{\,\prime}=1,2,3,\qquad
\psi_{\,\mu\mu^{\prime}}>0,\quad 
\qquad (\mu>\mu^{\,\prime}).
\label{psi-mu-mu-3-branches}
\eeq
Then the linearized equations are obtained as 
\beq
\dot{Z}_{\,1}
=-\,\underbrace{\psi_{\,0}\,\psi_{\,21}\psi_{\,31}}_{>\ 0}\, Z_{\,1},
\qquad
\dot{Z}_{\,2}
=\ \ \underbrace{\psi_{\,0}\,\psi_{\,21}\psi_{\,32}}_{>\ 0}\, Z_{\,2},
\qquad
\dot{Z}_{\,3}
=
-\,\underbrace{\psi_{\,0}\psi_{\,31}\psi_{\,32}}_{>\ 0}\, Z_{\,3},
\label{linearized-Z-3-branches}
\eeq
and 
\beqa
\dot{Y}_{\,1}
&=&-\psi_{\,0}\,\psi_{\,21}\psi_{\,31}\, Y_{\,1}
+(\,\psi_{\,0}^{\,\prime}\,\psi_{\,21}\,\psi_{\,31}
+\psi_{\,0}\,\psi_{\,21}^{\,\prime}\,\psi_{\,31}
+\psi_{\,0}\,\psi_{\,21}\,\psi_{\,31}^{\,\prime})\,Z_{\,1},
\non\\
\dot{Y}_{\,2}
&=&\ \psi_{\,0}\,\psi_{\,21}\psi_{\,32}\, Y_{\,2}
-(\,\psi_{\,0}^{\,\prime}\,\psi_{\,21}\,\psi_{\,32}
+\psi_{\,0}\,\psi_{\,21}^{\,\prime}\,\psi_{\,32}
+\psi_{\,0}\,\psi_{\,21}\,\psi_{\,32}^{\,\prime})\,Z_{\,2},
\non\\
\dot{Y}_{\,3}
&=&-\psi_{\,0}\,\psi_{\,31}\psi_{\,32}\, Y_{\,3}
+(\,\psi_{\,0}^{\,\prime}\,\psi_{\,31}\,\psi_{\,32}
+\psi_{\,0}\,\psi_{\,31}^{\,\prime}\,\psi_{\,32}
+\psi_{\,0}\,\psi_{\,31}\,\psi_{\,32}^{\,\prime})\,Z_{\,3}.
\non
\eeqa
To solve this linear system of equations, letting $c_{\,\mu}$ 
and $d_{\,\mu}$ be some constants, one can write the system as 
$$
\dot{Z}_{\,\mu}
=-\,c_{\,\mu} Z_{\,\mu},\quad
\dot{Y}_{\,\mu}
=-\,c_{\,\mu} Y_{\,\mu}+d_{\,\mu}Z_{\,\mu}.
$$
The solution of this system is 
$$
Z_{\,\mu}(t)
=Z_{\,\mu}(0)\,\e^{\,-c_{\mu}\,t},\quad
Y_{\,\mu}(t)
=(\,Y_{\,\mu}(0)+d_{\,\mu}Z_{\,\mu}(0)\, t\,)\,\e^{\,-c_{\mu}\,t}.
$$
From this and the inequalities $c_{\,1}>0$, $c_{\,3}>0$, $c_{\,2}<0$, 
one has that the fixed point sets 
$\phi\cA_{\,1}^{\,\cI^{+}}$ and $\phi\cA_{\,3}^{\,\cI^{+}}$ 
are linearly stable, and the
$\phi\cA_{\,2}^{\,\cI^{+}}$ is linearly unstable. 
Observe from
\fr{psi-mu-mu-3-branches} and
\fr{linearized-Z-3-branches} 
that the strength of instability/stability is large when
the value $\psi_{\,\mu\mu^{\,\prime}}$ is large. 
The condition when $\psi_{\,\mu\mu^{\,\prime}}(x)$ is large can be read off 
from Fig.\,\ref{Husimi-wave-front-picture}.
The values $\psi_{\,\mu\mu^{\prime}}$ are small near the critical
point, and  
they are  large far from the critical point.

\subsection{Two branch system}

We focus on the low temperature phase again, and
the following is about another contact Hamiltonian system.  

\begin{Theorem}
  \label{claim-cubic-contact-Hamiltonian2}
  (stable segment of the hysteresis curve in the symmetry broken phase).
  On the thermodynamic phase space
  for the Husimi-Temperley model, 
  choose a contact Hamiltonian $h$ as 
\beq
  h(x,z)
  = \psi_{\,0}(x)(z-\psi_{\,1}(x))(z-\psi_{\,2}(x)),
\label{wave-fronts-as-attractor-contact-hamiltonian2}
  \eeq
  where
  $\psi_{\,0}$  is an arbitrary function of $x$ such that  $\psi_{\,0}(x)>0$.
Then the following holds. 
  \begin{enumerate}
\item    
  The space $\phi\cA_{\,1}^{\,\cI^{\,+}}$ is asymptotically stable
  in $\cD_{\,1}^{\,+}$,  
  where  
  $$
  \cD_{\,1}^{\,+}
  =  \{\ (x,y,z)\ | \ x\in\cI^{\,+},\ z<\psi_{\,2}(x)\  
  \}.
  $$
  \end{enumerate}
\end{Theorem}
\begin{Proof}
  Our strategy for proving this is to show the existence of a
  Lyapunov function\,\cite{smale} for  
  the dynamical system, where this system is obtained from
  substituting the contact Hamiltonian 
\fr{wave-fronts-as-attractor-contact-hamiltonian2} into 
\fr{contact-hamiltonian-vector-field-component}. 
The details are as follows.

First, a point of departure for this proof is to express 
the explicit form of the
dynamical system written in terms of the coordinates $(x,y,z)$.  
From \fr{contact-hamiltonian-vector-field-component}
and \fr{wave-fronts-as-attractor-contact-hamiltonian2}, the
dynamical system is explicitly written as   
\beqa
\dot{x}&=&0,
\label{wave-fronts-as-attractor-x2}\\
\dot{y}
&=&-\,\frac{\dr \psi_{\,0}}{\dr x}(z-\psi_{\,1})(z-\psi_{\,2})
+\psi_{\,0}\left(y+\frac{\dr\psi_{\,1}}{\dr x}\right)(z-\psi_{\,2})
+\psi_{\,0}\left(y+\frac{\dr\psi_{\,2}}{\dr x}\right)(z-\psi_{\,1})
\label{wave-fronts-as-attractor-y2}\\
\dot{z}
&=&h
= \psi_{\,0}(x)(z-\psi_{\,1}(x))(z-\psi_{\,2}(x)).
\label{wave-fronts-as-attractor-z2}
\eeqa
The next step is to find fixed point sets. 
From 
$$
\dot{x}|_{\,\phi\cA_{\mu}^{\cI^{+}}}
=0,\quad
\dot{y}|_{\,\phi\cA_{\mu}^{\cI^{+}}}
=0,\quad
\dot{z}|_{\,\phi\cA_{\mu}^{\cI^{+}}}
=0,\qquad \mu=1,2
$$
one has that $\phi\cA_{\,\mu}^{\cI^{+}}\subset\cC$, $(\mu=1,2)$
forms a fixed point set for each $\mu$. 
Here a phase portrait of the dynamical system is roughly discussed.
It follows from 
\fr{wave-fronts-as-attractor-x2} that $x$ is constant in time, and
thus $\psi_{\,\mu}(x)$ does not depend on time.

Third, to prove the theorem, 
a Lyapunov function is constructed\,\cite{smale}.
Define the function $V_{\,1}$ in $\cD_{\,1}^{\,+}$ 
such that 
$$
V_{\,1}(x,z)
=\frac{1}{2}(z-\psi_{\,1}(x))^{\,2},\qquad (x,y,z)\in\cD_{\,1}^{\,+}.
$$
Then, it follows that 
$$
V_{\,1}(x,z)
\geq 0,\quad
\frac{\dr V_{\,1}}{\dr t}
=(z-\psi_{\,1})h(x,z)
=\psi_{\,0}(z-\psi_{\,1})^{\,2}(z-\psi_{\,2})
\leq 0,\qquad (x,y,z)\in\cD_{\,1}^{\,+},
$$
where the equality holds on the fixed point set 
$\phi\cA_{\,1}^{\,\cI^{\,+}}$.
Hence $V_{\,1}$ is a Lyapunov function in $\cD_{\,1}^{\,+}$. 
According to the theorem of Lyapunov, one completes the proof.
\qed
\end{Proof}

Theorem\,\ref{claim-cubic-contact-Hamiltonian2} shows that  
the proposed contact Hamiltonian vector field is such that
a segment of the projected Legendre submanifold is stable 
in a region of a contact manifold. 
The global behavior for $z$ is understood from
Fig.\,\ref{Husimi-legndre-picture-flow-z2}.
One then deduces from Fig.\,\ref{Husimi-legndre-picture-flow-z2}
that, given $x$, 
$\lim_{t\to\infty}z(t)=\psi_{\,1}(x)$ in $\cD_{\,1}^{\,+}$.

\begin{figure}[htb]
\begin{picture}(120,70)
\unitlength 1mm
\put(47,26){$\dot{z}$}
\put(95,10){$z$}
\put(30,18){$\dot{z}=h(x,z)$}
\put(51,11){\line(0,1){15}}
\put(51,11){\line(1,0){40}}
\put(47,10){$0$}
\qbezier(50,20)(60,0)(75,10)
\qbezier(75,10)(80,15)(85,20)
\put(51,3){$z=\psi_{\,1}$}
\put(65,17){$z=\psi_{\,2}(x)$}
\put(56,5){\vector(0,1){5}}
\put(76.4,15.7){\vector(0,-1){5}}
\linethickness{0.5mm}
\end{picture}
\caption{Phase space of
  the dynamical system consisting of $\dot{z}=h(x,z)$ and $\dot{x}=0$
  (\fr{wave-fronts-as-attractor-z2} and 
  \fr{wave-fronts-as-attractor-x2}, respectively).
 From $\dot{z}=h$, it follows that the zeros of
  $h$ are the fixed points.
 From $h$ in \fr{wave-fronts-as-attractor-contact-hamiltonian2}
  its zeros are the set $z=\psi_{\,1}(x)$ and the set $z=\psi_{\,2}(x)$. 
  In addition, from the sign of $h$, it follows that 
  the set  $z=\psi_{\,1}(x)$  
  is stable in $\cD_{\,1}^{\,+}$.
  }
\label{Husimi-legndre-picture-flow-z2}
\end{figure}

Physically, the contact Hamiltonian vector field
with \fr{wave-fronts-as-attractor-contact-hamiltonian2} 
expresses the dynamical
process departing from initial states in $\cD_{\,1}^{\,+}$ 
to the most stable equilibrium ones.  

To grasp local flow around the fixed point sets, the 
integral curves of the linearized equations are shown below.
For the point  $(y_{\mu},z_{\mu})=(- \psi_{\,\mu}^{\,\prime},\psi_{\,\mu})$,
introduce $Y_{\,\mu}$ and $Z_{\,\mu}$
such that 
$$
y(t)
=- \psi_{\,\mu}^{\,\prime}(x)+Y_{\,\mu}(t),\quad
z(t)
=\psi_{\,\mu}(x)+Z_{\,\mu}(t),\quad\mbox{where}\quad
\psi_{\,\mu}^{\,\prime}(x)
:=\frac{\dr \psi_{\,\mu}}{\dr x}(x),
\quad\mu=1,2
$$
which yield linearized equations. For ease of notation, introduce 
\beq
\psi_{\,21}(x)
=\psi_{\,2}(x)-\psi_{\,1}(x)\ 
>0, \quad
\label{linearized-Z-2-branches}
\eeq
for each point $x$. 
Then the linearized equations are obtained as 
\beqa
\dot{Z}_{\,1}
&=&-\,\underbrace{\psi_{\,0}\,\psi_{\,21}}_{>\ 0}\, Z_{\,1},
\qquad
\dot{Z}_{\,2}
=\ \ \underbrace{\psi_{\,0}\,\psi_{\,21}}_{>\ 0}\, Z_{\,2},
\non\\
\dot{Y}_{\,1}
&=&-\psi_{\,0}\,\psi_{\,21}\, Y_{\,1}
+(\,\psi_{\,0}^{\,\prime}\,\psi_{\,21}
+\psi_{\,0}\,\psi_{\,21}^{\,\prime}\,)\,Z_{\,1},
\qquad
\dot{Y}_{\,2}
=\ \psi_{\,0}\,\psi_{\,21}\, Y_{\,2}
-(\psi_{\,0}^{\,\prime}\,\psi_{\,21}
+\psi_{\,0}\,\psi_{\,21}^{\,\prime}\,)\,Z_{\,2}, 
\non
\eeqa
where $\psi_{\,21}^{\,\prime}(x)
=\dr\psi_{\,21}/\dr x$ and 
$\psi_{\,0}^{\,\prime}=\dr\psi_{\,0}/\dr x$ that are constants in time.  
To solve this linear system of equations, letting $c_{\,\mu}$ 
and $d_{\,\mu}$ be the constants such that
$$
c_{\,1}
=\psi_{\,0}\,\psi_{\,21},\quad
c_{\,2}
=-\,c_{\,1},\qquad
d_{\,1}
=\psi_{\,0}^{\,\prime}\,\psi_{\,21}+\psi_{\,0}\,\psi_{\,21}^{\,\prime},
\quad
d_{\,2}
=-d_{\,1},
$$
one can write 
$$
\dot{Z}_{\,\mu}
=-\,c_{\,\mu} Z_{\,\mu},\quad
\dot{Y}_{\,\mu}
=-\,c_{\,\mu} Y_{\,\mu}+d_{\,\mu}Z_{\,\mu},\quad
\mu=1,2.
$$
The solution of this system is 
$$
Z_{\,\mu}(t)
=Z_{\,\mu}(0)\,\e^{\,-c_{\mu}\,t},\quad
Y_{\,\mu}(t)
=(\,Y_{\,\mu}(0)+d_{\,\mu}Z_{\,\mu}(0)\, t\,)\,\e^{\,- c_{\mu}\,t}.
$$
From this, the inequalities $c_{\,1}>0$ and $c_{\,2}<0$, 
one has that the fixed point set  
$\phi\cA_{\,1}^{\cI^{\,+}}$  
is linearly stable, and that the $\phi\cA_{\,2}^{\cI^{\,+}}$ 
linearly unstable. 
Observe from
\fr{linearized-Z-2-branches} 
that the strength of instability/stability is large when the value  
$c_{\,1}=\psi_{\,0}\,\psi_{\,21}$ is large. 
The condition when $\psi_{\,21}(x)$ is large can be read off 
from Fig.\,\ref{Husimi-partial-hysteresis1-picture}.
The value $\psi_{\,21}(x)$ is small near the critical point, and  
it is large far from the critical point. 

So far the phase space  $\cD_{\,1}^{\,+}$ of the dynamical system is focused,  
and then a similar claim can be stated for the
region $\cD_{\,1}^{\,-}$, where 
$\cD_{\,1}^{\,-}$ is defined with some $\cI^{\,-}\subset\mbbR_{<0}$.
To state such a claim, recall the following:
$$
\cD_{\,1}^{\,+}
=\{(x,y,z)\ |\,x\in\cI^{\,+},z<\psi_{\,2}(x)\},\quad
\cI^{\,+}
=\{\, x\in\mbbR_{\,>0}\,|\,\psi_{\,1}(x)<\psi_{\,2}(x)\,\} \subset\mbbR,
$$
and 
$$
\cD_{\,1}^{\,-}
=\{(x,y,z)\ |\,x\in\cI^{\,-},z<\psi_{\,2}(x)\},\quad
\cI^{\,-}
=\{\, x\in\mbbR_{\,<0}\,|\,\psi_{\,1}(x)<\psi_{\,2}(x)\,\} \subset\mbbR.
$$

By combining these and refining it, one has the following.
\begin{Corollary}
  \label{fact-combining-the-contact-Hamiltonian-system}
  (reconstruction of the stability of the hysteresis and
  pseudo-free energy curves as the Legendre submanifold).  
  Consider the system shown in Fig.\,\ref{Husimi-Legendre-hysteresis-picture}
  with the curve $\ol{\mathrm{ii-iv}}$ being removed. 
  In the joined region 
  $\cD_{\,1}^{\,+}\cup \cD_{\,1}^{\,-}$ in the contact manifold $\cC$,  
  one has the contact Hamiltonian vector fields where the undirected curves
  $\ol{\mathrm{0-i}}$ and  $\ol{\mathrm{v-vi}}$ are stable fixed point sets.
\end{Corollary}  
Notice that the  $(y,z)$-plane at $x=0$  
has been removed from $\cC$ in Corollary\,\ref{fact-combining-the-contact-Hamiltonian-system}. 
On this removed plane, the
  double-valued function becomes a single valued function, and thus  
  the present contact Hamiltonian is not relevant. 
  The projected contact Hamiltonian vector fields 
stated  in
Corollary\,\ref{fact-combining-the-contact-Hamiltonian-system} can 
be shown, and they are the same as 
Fig.\,\ref{Husimi-partial-hysteresis1-vector-field-picture}. 
From this corollary, one has again Remark\,\ref{remark-asymptotic-phase-portrait}.



\end{document}